\documentclass[12pt]{amsart}
\usepackage{amsfonts,amssymb}
\RequirePackage{ifpdf}
\usepackage{amsmath}
\usepackage{color}
\usepackage{pstricks,pst-node}
\usepackage{pst-plot}
\def\appendix#1{
\addtocounter{section}{1} \setcounter{equation}{0}
\renewcommand{\thesection}{\Alph{section}}
\section*{Appendix \thesection\protect\indent\quad
#1}
}
\renewcommand{\theequation}{\thesection.\arabic{equation}}

\catcode`\@=11
\def\marginnote#1{}

\newcount\hour
\newcount\minute
\newtoks\amorpm
\hour=\time\divide\hour by60 \minute=\time{\multiply\hour by60
\global\advance\minute by-\hour}
\edef\standardtime{{\ifnum\hour<12 \global\amorpm={am}%
        \else\global\amorpm={pm}\advance\hour by-12 \fi
        \ifnum\hour=0 \hour=12 \fi
        \number\hour:\ifnum\minute<10 0\fi\number\minute\the\amorpm}}
\edef\militarytime{\number\hour:\ifnum\minute<100\fi\number\minute}

\newcommand{\tcr}{\textcolor{red}}

\def\draftlabel#1{{\@bsphack\if@filesw {\let\thepage\relax
      \xdef\@gtempa{\write\@auxout{\string
          \newlabel{#1}{{\@currentlabel}{\thepage}}}}}\@gtempa \if@nobreak
    \ifvmode\nobreak\fi\fi\fi\@esphack} \gdef\@eqnlabel{#1}}
    \def\@eqnlabel{}
\def\@vacuum{}
\def\draftmarginnote#1{\marginpar{\raggedright\scriptsize\tt#1}}

\def\draft{
  \oddsidemargin -.5truein
  \def\@oddfoot{\footnotesize \sl preliminary draft \hfil
    \rm\thepage\hfil\sl\today\quad\militarytime}
  \let\@evenfoot\@oddfoot \overfullrule 3pt
    \let\label=\draftlabel
    \let\marginnote=\draftmarginnote
  \def\@eqnnum{(\theequation)\rlap{\kern\marginparsep\tt\@eqnlabel}%
    \global\let\@eqnlabel\@vacuum}

  }

\voffset= - 0.5in \hoffset= - 1.0in

\def\be{\begin{equation}}
\def\ee{\end{equation}}
\def\bea{\begin{eqnarray}}
\def\eea{\end{eqnarray}}
\def\<{\langle}
\def\>{\rangle}

\def\ocomma{{\phantom{\Bigm|}^{\phantom {X}}_{\raise-1.5pt\hbox{,}}\!\!\!\!\!\!\otimes}}

\newtheorem{theorem}{Theorem}[section]
\newtheorem{lemma}[theorem]{Lemma}
\newtheorem{proposition}[theorem]{Proposition}
\newtheorem{corollary}[theorem]{Corollary}
\theoremstyle{definition}

\newtheorem{remark}[theorem]{Remark}

\allowdisplaybreaks

\long\def\rem#1{}

\begin{document}

\title[Fool's crowns, trumpets, and Schwarzian]
{Fool's crowns, trumpets, and Schwarzian}
\author{Leonid O. Chekhov$^{\ast}$}\thanks{$^{\ast}$Michigan State University, East Lansing, USA, and Steklov Mathematical
Institute of Russian Academy of Sciences, Moscow, Russia. Email: chekhov@msu.edu.}

\begin{abstract}
For a Riemann surface with holes, we propose a variant of the action on a circum\-ference-$P$ boundary component with $n$ bordered cusps attached (a ``fool's crown'') that is decoration-invariant and generates finite volumes $V^{\text{crown}}_{n,P}$ of the corresponding moduli spaces when integrated against the volume form obtained by inverting the Fenchel--Nielsen (Goldman) Poisson brackets for a special set of decoration-invariant combinations of Penner's $\lambda$ lengths. In the limit as $n\to\infty$, the integrals transform into a functional integral with the measure given by the integral over $C^1$ of the action $A_1^{(0)}-\frac12 S[\psi,t]+\frac 12 (\psi')^2$. Here $A_1^{(0)}\sim \int \log \psi' \frac {dx}x$ is the disc amplitude, $S[\psi,t]$ is the Schwarzian, and the derivative $\psi'$ is related to the limiting density of orthogonal projections of bordered cusps to the hole perimeter. We derive the Fenchel--Nielsen symplectic form in the continuum limit and show that it coincides with the one obtained by Alekseev and Meinrenken. We also discuss the volumes of moduli spaces for a disc with $n$ bordered cusps. 
\end{abstract}

\maketitle

\section{Introduction}\label{s:intro}
\setcounter{equation}{0}

Theories describing boundaries of Riemann surfaces got a boost recently from a novel insight into Jackiw--Teitelboim (JT) theory of gravity related to its supersymmetric generalizations (due to Stanford and Witten (SW) \cite{SW}, Norbury \cite{Nor}, end others) Description of moduli spaces of Riemann surfaces with holes has a long and successful story, both in complex-analytic uniformization (due to Strebel, Kontsevich, Witten) and in  real-analytic (hyperbolic) Poincar\'e uniformization (due to Penner, Thurston, V.V.Fock and many others). In all these approaches dated back to XX century, structures on the boundary itself (trumpets, in SW terminology) were mostly overlooked in mathematics.\footnote{Worth mentioning is however an acitve development of topological models related to Wess--Zumino--Novikov--Witten theory \cite{Wit1} and to Virasoro group action on the boundary \cite{Al-Shat}.} A possible reason was that introducing these, generally infinitely-dimensional, structures does not enlarge sufficiently the mapping-class group action (besides introducing a one new twist along the hole perimeter), so the volumes of related moduli spaces turn out to be infinite. 

On the other hand, a description that involves nontrivial structures on the boundary of a hole was developed in \cite{ChMaz2}, where this description was originated from confluences of holes in the standard pattern of Riemann surfaces with holes. These structures, called \emph{bordered cusps} (to distinguish them from standard cusps, which are just holes of zero circumference) correspond topologically to marked points on hole boundaries, and we have the corresponding Teichm\"uller spaces $\mathfrak T_{g,s,n}$, where $g$, $s$, and $n$ denote correspondingly the genus, number of holes $s>0$, and number $n$ of bordered cusps on hole boundaries. Rigorously speaking, instead of $n$ we have to indicate the corresponding numbers for each boundary component, $n_1,\dots, n_s$ with $n_1+n_2+\dots+n_s=n$, because the number of marked points on each boundary component is fixed and cannot be changed by a mapping-class group action. Note that these bordered cusps have to be decorated with horocycles, and admit a topological interpretation as marked points on boundaries of holes of $\Sigma_{g,s}$; a quantitative description of surfaces with marked points on  boundary components was given from various points of view by Fock and Goncharov \cite{FG1}, Musiker, Schiffler and Williams \cite{MSW1}, \cite{MSW2}, \cite{MW}, and S. Fomin, M. Shapiro, and D. Thurston \cite{FST}, \cite{FT}.

Volumes of hyperbolic moduli spaces, known as Mirzakhani's volumes $V_g(P_1,\dots,P_s)$ \cite{Mir06} with $P_i$ the circumferences of holes, were obtained by factoring Teichm\"uller spaces ${\mathfrak T}_{g,s}$ by the action of the mapping-class group. With fixed $P_i$, these volumes are finite and are polynomial in $P_i$. The recurrence relations derived in \cite{Mir07} for these volumes were shown by Mulase, Safnuk \cite{MS06} and Do and Norbury \cite{DoN06} to satisfy Virasoro algebra relations, which eventually resulted in the construction of topological recursion \cite{CEO} model by Eynard and Orantin \cite{EO07} governing the generating function for these volumes. 

Since adding new bordered cusps does not affect the modular group, if we address the problem in the same way, we obtain infinite volumes (if at least one of $n_i>1$). It is thus essential to introduce a regularization, or action functional. Different choices of this action functional result in very different answers for the corresponding volumes. We advocate a choice (\ref{act1}) for this action based on its invariance and analyticity properties.

In each homotopy class of curves starting and terminating at bordered cusps (can be the same cusp) we have a unique geodesic curve $\mathfrak a$ of infinite length, which we call an \emph{arc}. The Penner \cite{Penn1} $\lambda$-length of this arc is, by definition, $\lambda=e^{\ell_{\mathfrak a}/2}$, where $\ell_{\mathfrak a}$ is the signed hyperbolic length of the segment of this geodesic confined between two horocycles decorating the cusps in which this geodesic starts and terminate. These lambda lengths satisfy the celebrated classical ``Ptolemy'' relation discovered by Penner \cite{Penn1} and are subject to Poisson relations virtually identical to the standard Goldman brackets \cite{Gold} for \emph{geodesic functions} (doubled hyperbolic cosines of half-lengths of closed geodesics). 

\subsection{Goldman brackets for arcs}\label{ss:Goldman}

Decoration of bordered cusps with horocycles allows constructing an extension of Poisson structure on the set of shear coordinates by adding a ``hybrid,'' or ``extended'' shear coordinates depending on decoration (one such coordinate for each bordered cusp \cite{ChMaz2}). These ``extended'' shear coordinates have nontrivial Poisson brackets with other coordinates.

The Goldman Poisson bracket for arcs ($\lambda$-lengths) was introduced in \cite{ChMaz2}; we have to supply the standard Goldman bracket with the case of arcs incident to the same bordered cusp. Note that the Poisson relations for $\lambda$-lengths of arcs cannot be defined unless we have at least one bordered cusp.

When two arcs, say $\mathfrak a_{1,3}$ and $\mathfrak a_{2,4}$ intersect at  an internal point of the surface, we set (see Fig.~\ref{fi:Goldman1})
\be
\{\lambda_{1,3},\lambda_{2,4}\}=\frac 12 \lambda_{1,4}\lambda_{2,3} -\frac 12 \lambda_{1,2}\lambda_{3,4},
\label{Gold-inner}
\ee

\begin{figure}[tb]
\begin{pspicture}(-4,-1.5)(4,1.5){
\rput(-3,0){
\psclip{\pscircle[linewidth=1.5pt, linestyle=dashed](0,0){1}}
\rput(0,0){\psline[linewidth=1.5pt,linecolor=red, linestyle=dashed](1,-1)(-1,1)}
\rput(0,0){\psline[linewidth=3pt,linecolor=white](-1,-1)(1,1)}
\rput(0,0){\psline[linewidth=1.5pt,linecolor=blue,  linestyle=dashed](-1,-1)(1,1)}
\endpsclip
\rput(0,-1.2){\makebox(0,0)[ct]{$\{\lambda_{1,3}\,\lambda_{2,4}\}$}}
\rput(0.8,0.8){\makebox(0,0)[lb]{$3$}}
\rput(0.8,-0.8){\makebox(0,0)[lt]{$4$}}
\rput(-0.8,0.8){\makebox(0,0)[rb]{$2$}}
\rput(-0.8,-0.8){\makebox(0,0)[rt]{$1$}}
}
\rput(0,0){
\psclip{\pscircle[linewidth=1.5pt, linestyle=dashed](0,0){1}}
\rput(0,-1.4){\psarc[linewidth=1.5pt,linecolor=green, linestyle=dashed](0,0){1}{45}{135}}
\rput(0,1.4){\psarc[linewidth=1.5pt,linecolor=green, linestyle=dashed](0,0){1}{225}{315}}
\endpsclip
\rput(0.8,0.8){\makebox(0,0)[lb]{$3$}}
\rput(0.8,-0.8){\makebox(0,0)[lt]{$4$}}
\rput(-0.8,0.8){\makebox(0,0)[rb]{$2$}}
\rput(-0.8,-0.8){\makebox(0,0)[rt]{$1$}}
\rput(0,-1.2){\makebox(0,0)[ct]{$\lambda_{1,4}\,\lambda_{2,3}$}}
\rput(-1.2,0){\makebox(0,0)[rc]{$\dfrac12$}}
}
\rput(3,0){
\psclip{\pscircle[linewidth=1.5pt, linestyle=dashed](0,0){1}}
\rput(-1.4,0){\psarc[linewidth=1.5pt,linecolor=green, linestyle=dashed](0,0){1}{-45}{45}}
\rput(1.4,0){\psarc[linewidth=1.5pt,linecolor=green, linestyle=dashed](0,0){1}{135}{225}}
\endpsclip
\rput(0.8,0.8){\makebox(0,0)[lb]{$3$}}
\rput(0.8,-0.8){\makebox(0,0)[lt]{$4$}}
\rput(-0.8,0.8){\makebox(0,0)[rb]{$2$}}
\rput(-0.8,-0.8){\makebox(0,0)[rt]{$1$}}
\rput(0,-1.2){\makebox(0,0)[ct]{$\lambda_{1,2},\lambda_{3,4}$}}
\rput(-1.2,0){\makebox(0,0)[rc]{$\dfrac12$}}
}
\rput(-1.7,0){
\rput(0,0){\makebox(0,0){$=$}}}
\rput(1.3,0){
\rput(0,0){\makebox(0,0){$-$}}}
}
\end{pspicture}
\caption{\small The ``elementary'' Poisson bracket (the Goldman bracket) $\{\lambda_{1,3},\lambda_{2,4}\}$ (\ref{Gold-inner}) between two $\lambda$-lengths
of the corresponding arcs intersecting at a point inside a Riemann surface. We also have the classical skein relation (the Penner's ``Ptolemy'' relation, not depicted) $\lambda_{1,3}\,\lambda_{2,4}=\lambda_{1,4}\,\lambda_{2,3}+\lambda_{1,2}\,\lambda_{3,4}$.}
\label{fi:Goldman1}
\end{figure}
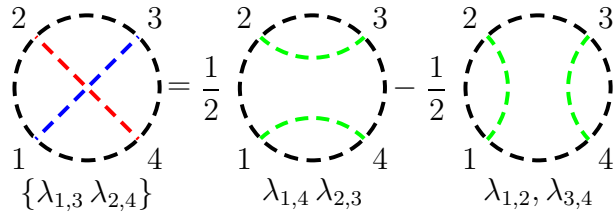

\begin{figure}[tb]
\begin{pspicture}(-4,-1.5)(4,1.5){
\rput(-5,0){
\psclip{\pscircle[linewidth=1.5pt, linestyle=dashed](0,0){1}}
\rput(-1,2){\psarc[linewidth=1.5pt,linecolor=blue, linestyle=dashed](0,0){2}{270}{360}}
\rput(-1,-2){\psarc[linewidth=1.5pt,linecolor=red, linestyle=dashed](0,0){2}{0}{90}}
\rput(-0.7,0){\pscircle[linewidth=0pt, linecolor=white,fillstyle=solid, fillcolor=white](0,0){0.3}}
\rput(-1,1){\pscircle[linewidth=1pt,linecolor=green, linestyle=dashed, fillstyle=solid, fillcolor=lightgray](0,0){1}}
\rput(-1,-1){\pscircle[linewidth=1pt,linecolor=green, linestyle=dashed, fillstyle=solid, fillcolor=lightgray](0,0){1}}
\psclip{\pscircle[linestyle=dashed,linewidth=1.5pt](-0.7,0){0.3}}
\rput(-1,2){\psarc[linewidth=.5pt,linecolor=blue](0,0){2}{270}{320}}
\rput(-1,-2){\psarc[linewidth=.5pt,linecolor=red](0,0){2}{60}{90}}
\endpsclip
\psarc[linewidth=1pt]{<-}(-0.5,0){.5}{-80}{80}
\endpsclip
\rput(0.8,0.8){\makebox(0,0)[lb]{$\mathfrak a_1$}}
\rput(0.8,-0.8){\makebox(0,0)[lt]{$\mathfrak a_2$}}
%
\rput(0,-1.2){\makebox(0,0)[ct]{$\{\lambda_1,\lambda_2\}$}}
}
\rput(-2,0){
\psclip{\pscircle[linewidth=1.5pt, linestyle=dashed](0,0){1}}
\rput(-1,2){\psarc[linewidth=1.5pt,linecolor=blue, linestyle=dashed](0,0){2}{270}{360}}
\rput(-1,-2){\psarc[linewidth=1.5pt,linecolor=red, linestyle=dashed](0,0){2}{0}{90}}
\rput(-0.7,0){\pscircle[linewidth=0pt, linecolor=white,fillstyle=solid, fillcolor=white](0,0){0.3}}
\rput(-1,1){\pscircle[linewidth=1pt,linecolor=green, linestyle=dashed, fillstyle=solid, fillcolor=lightgray](0,0){1}}
\rput(-1,-1){\pscircle[linewidth=1pt,linecolor=green, linestyle=dashed, fillstyle=solid, fillcolor=lightgray](0,0){1}}
\psclip{\pscircle[linestyle=dashed,linewidth=1.5pt](-0.7,0){0.3}}
\rput(-1,2){\psarc[linewidth=.5pt,linecolor=blue](0,0){2}{270}{320}}
\rput(-1,-2){\psarc[linewidth=.5pt,linecolor=red](0,0){2}{60}{90}}
\endpsclip
\endpsclip
\rput(0.8,0.8){\makebox(0,0)[lb]{$\mathfrak a_1$}}
\rput(0.8,-0.8){\makebox(0,0)[lt]{$\mathfrak a_2$}}
%
\rput(0,-1.2){\makebox(0,0)[ct]{$\lambda_1\,\lambda_2$}}
\rput(-1.2,0){\makebox(0,0)[rc]{$=\dfrac 14$}}
}
}
\rput(1.5,0){
\psclip{\pscircle[linewidth=1.5pt, linestyle=dashed](0,0){1}}
\rput(0,0){\psbezier[linewidth=1.5pt,linecolor=red, linestyle=dashed](-1,0)(-0.3,0)(0.5,0.5)(0.5,0)
\psbezier[linewidth=1.5pt,linecolor=red, linestyle=dashed](-1,0)(-0.3,0)(0.5,-0.5)(0.5,0)
}
\rput(-0.7,0){\pscircle[linewidth=0pt, linecolor=white,fillstyle=solid, fillcolor=white](0,0){0.3}}
\rput(-1,1){\pscircle[linewidth=1pt,linecolor=green, linestyle=dashed, fillstyle=solid, fillcolor=lightgray](0,0){1}}
\rput(-1,-1){\pscircle[linewidth=1pt,linecolor=green, linestyle=dashed, fillstyle=solid, fillcolor=lightgray](0,0){1}}
\psclip{\pscircle[linestyle=dashed,linewidth=1.5pt](-0.7,0){0.3}}
\rput(-1,2){\psarc[linewidth=.5pt,linecolor=red](0,0){2}{270}{320}}
\rput(-1,-2){\psarc[linewidth=.5pt,linecolor=red](0,0){2}{60}{90}}
\endpsclip
\endpsclip
\rput(1.2,0){\makebox(0,0)[lc]{$=0$}}
}
\rput(5,0){
\psclip{\pscircle[linewidth=1.5pt, linestyle=dashed](0,0){1}}
%
\rput(-0.7,0){\pscircle[linewidth=0pt, linecolor=white,fillstyle=solid, fillcolor=white](0,0){0.3}}
\rput(-1,1){\pscircle[linewidth=1pt,linecolor=green, linestyle=dashed, fillstyle=solid, fillcolor=lightgray](0,0){1}}
\rput(-1,-1){\pscircle[linewidth=1pt,linecolor=green, linestyle=dashed, fillstyle=solid, fillcolor=lightgray](0,0){1}}
\rput(0,0){\pscircle[linestyle=dashed,linewidth=1.5pt,linecolor=red](0.3,0){0.5}}
\psclip{\pscircle[linestyle=dashed,linewidth=1.5pt](-0.7,0){0.3}}
\endpsclip
\endpsclip
\rput(1.2,0){\makebox(0,0)[lc]{$=-2$}}
}
\end{pspicture}
\caption{\small The ``elementary'' Poisson bracket (the Goldman bracket) $\{\lambda_1,\lambda_2\}$ (\ref{Goldman2}) between 
lambda lengths of the corresponding arcs ${\mathfrak a}_1$ and ${\mathfrak a}_2$ coming to the same bordered cusp
of a Riemann surface; we also indicate that the loop contractible to a bordered cusp is equal zero thus killing the whole lamination that contains such a loop; an empty contractible loop gives the factor $-2$.}
\label{fi:Goldman2}
\end{figure}
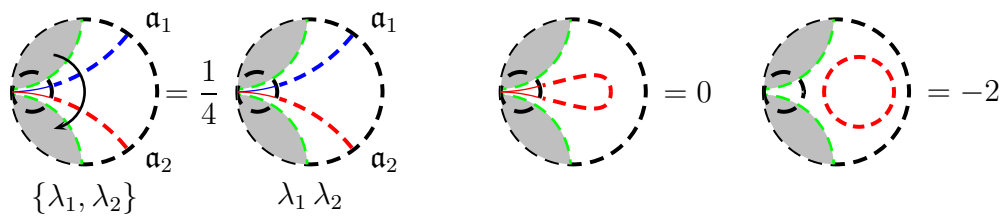

When two arcs meet at the same bordered cusp, we have (see Fig.~\ref{fi:Goldman2})
\be
\{\lambda_{1},\lambda_{2}\}= \frac 14 \lambda_{1}\lambda_{2} \quad  \hbox{if\ } \mathfrak a_1\ \hbox{is to the right of}\ \mathfrak a_2
\label{Goldman2}
\ee
where the sign is plus if the arc $\mathfrak a_1$ lies to the right from the arc $\mathfrak a_2$ when looking ``from inside'' the Riemann surface. In general, two arcs may intersect inside the surface and share the same bordered cusps at their ends; their Poisson bracket is then to be evaluated using the Leibnitz rule: we implement the Goldman relation at every crossing and at every bordered cusp shared by these arcs and apply the Ptolemy relations at all remaining inner crossings.

\subsection{Poisson and symplectic structures in shear coordinates}\label{ss:Poisson}

Another powerful approach to the description of Teichm\"uller spaces ${\mathfrak T}_{g,s}$ is due to ideal-triangle decomposition of $\Sigma_{g,s}$ and related Thurston's shear coordinates $Z_\alpha$ \cite{ThSh}  for surfaces with punctures generalized by V.V.Fock \cite{Fock1} to surfaces with holes. A log-canonical mapping-class group invariant Poisson structure was introduced by Fock on the set of shear coordinates in \cite{F97}. This structure was quantized in \cite{ChF1}.

An explicit combinatorial construction of the corresponding classical geodesic  functions in terms of shear coordinates of decorated Teichm\"uller spaces for Riemann surfaces with $s>0$ holes was proposed in \cite{ChF2}: it was shown there that geodesic  functions of all closed geodesics are Laurent polynomials of exponentiated halves of shear coordinates with positive integer coefficients; these polynomials were obtained using a fat-graph description dual to the ideal-triangle decomposition of $\Sigma_{g,s}$;  these results were extended to orbifold Riemann surfaces in  \cite{ChSh} where generalized cluster transformations (cluster algebras with coefficients) were introduced. Note that shear coordinates can be identified with the $Y$-type cluster variables \cite{FZ},~\cite{FZ2}, and mapping-class morphisms can be identified with cluster mutations. Extended shear coordinates are then identified with frozen cluster variables.

One of the most attractive properties of the fat graph description is a very simple Poisson algebra on the set of shear coordinates \cite{F97}: let $Y_k$, $k=1,2,3 \mod 3$, denote  $Z$-variables of cyclically ordered edges incident to a three-valent vertex (see Sec.~\ref{s:fat} for examples). The Poisson (Weil--Petersson) bi-vector field  is then
\be
\label{WP-PB}
w_{\text{WP}}:=\sum_{{\hbox{\small 3-valent} \atop \hbox{\small vertices} }}
\,\sum_{k=1}^{3} \partial_{Y_k}\wedge \partial_{Y_{k+1}},
\ee
and this bi-vector has the same form in the presence of extended shear coordinates.

\begin{theorem}\label{th-WP}\cite{ChF2} \cite{ChMaz2}
The bracket \eqref{WP-PB}, where $Y_k$ incorporate both the standard and extended shear coordinates,  induces the {\em Goldman
bracket}  \cite{Gold} (see Fig.~\ref{fi:Goldman1}) on the set of geodesic functions and on the set of $\lambda$-lengths of arcs.
\end{theorem}

The center of the Poisson algebra {\rm(\ref{WP-PB})} is generated by sums of $Y_\alpha$ (taken with multiplicities) of all edges incident to a given hole, so, irrespectively on the number of bordered cusps, we have exactly $s$ independent central elements.

Whereas no well-defined Poisson bracket exists for $\lambda$-lengths in the absence of bordered cusps, a mapping-class group invariant  {\em symplectic structure} on the set of $\lambda$-lengths of arcs was introduced by Penner \cite{Penn92} for $\mathfrak T_{g,s}$:
\be
\label{WP-SS}
\Omega_{\text{WP}}:=\sum_{{\hbox{\small ideal} \atop \hbox{\small triangles} }}
\,\sum_{k=1}^{3}d\log \lambda_k\wedge d\log \lambda_{k+1}.
\ee
This symplectic structure can be easily extended to the case of $\mathfrak T_{g,s,n}$.

In the case of $\Sigma_{g,s,n}$ with $n>0$, the bracket (\ref{WP-PB}) induces homogeneous Poisson relations on the set of $\lambda$-lengths of arcs from the same ideal-triangle decomposition amounting to Berenstein--Zelevinsky quantum cluster algebras \cite{BerZel}; this Poisson structure was shown  \cite{Ch20} to be inverse to the symplectic structure (\ref{WP-SS}).

Quantum bordered Riemann surfaces $\Sigma_{g,s,n}$ and the corresponding quantum Teichm\"uller spaces $\mathfrak T^\hbar_{g,s,n}$ were constructed in \cite{ChMaz2}. Having just one bordered cusp on one of boundary components enables constructing an ideal-triangle decomposition of such a surface $\Sigma_{g,s,n}$ in which all arcs start and terminate at bordered cusps; in the dual fat graph $\Gamma_{g,s,n}$, all holes without cusps then have to be confined in monogons (loops), and admissible mapping-class-group morphisms amount to generalized cluster mutations \cite{ChSh} preserving this condition. For such ideal-triangle decompositions of $\Sigma_{g,s,n}$ we have a bijection between the extended shear coordinates and lambda lengths that enables one to determine Poisson and quantum relations for lambda lengths; they appear to be \cite{ChMaz2} correspondingly Poisson and quantum cluster algebras by Berenstein and Zelevinsky \cite{BerZel}. On the other hand, a mapping-class group invariant 2-form (\ref{WP-SS}) on the set of lambda-lengths generates the invariant 2-form on the set of extended shear coordinates \cite{BK}, \cite{Ch20}.

\subsection{The fool's crown}\label{ss:fool}

We now introduce the main object of interest in this paper, namely, a boundary component with $n$ bordered cusps, which, by its appearance, resembles a ``fool's crown." (the term ``crown'' for this structure was used by the author in \cite{Ch20-2}).

We enumerate the cusps from $1$ to $n(=0)$ and let $\lambda_{i,j}$ denote the $\lambda$-length of the arc (without self-intersections) running in positive direction between $i$th and $j$th cusps. (Note that, generally, $\lambda_{i,j}\ne \lambda_{j,i}$.) For $\lambda$-lengths of curves joining the cusps in the crown in the right side of Fig.~\ref{fi:n-bcsps}, the Poisson algebra reads \cite{Ch20-2}
\begin{align*}
\{\lambda_{0,i},\lambda_{0,j}\}&=\frac14 \lambda_{0,i}\lambda_{0,j}, \quad  \{\lambda_{i,0},\lambda_{j,0}\}=\frac14 \lambda_{i,0}\lambda_{j,0}\quad 1\le i<j\le n-1; \\
\{\lambda_{0,j},\lambda_{i,0}\}&=\frac 34 \lambda_{0,j}\lambda_{i,0} - \lambda_{0,i},\lambda_{j,0}, \quad 1\le i<j\le n-1;\\
\{\lambda_{0,i},\lambda_{j,0}\}&=\frac 14 \lambda_{0,i}\lambda_{j,0}, \quad 1\le i<j\le n-1;\\
\{\lambda_{0,i},\lambda_{0,0}\}&=\frac 12 \lambda_{0,i}\lambda_{0,0},\quad \{\lambda_{i,0},\lambda_{0,0}\}=-\frac 12 \lambda_{i,0}\lambda_{0,0},\quad 1\le i\le n-1;\\
\{\lambda_{0,i},\lambda_{B}\}&=\frac 12 \lambda_{0,i}\lambda_{B},\quad \{\lambda_{i,0},\lambda_{B}\}=-\frac 12 \lambda_{i,0}\lambda_{B},\quad 1\le i\le n-1;
\quad \{\lambda_{0,i},\lambda_{i,0}\}=0.
\end{align*}
where in the second line we have used the classical skein relation $\lambda_{i,0}\lambda_{0,j}=\lambda_{i,j}\lambda_{0,0}+\lambda_{j,0}\lambda_{0,i}$ satisfied for $i<j$. These brackets induce the Poisson bracket (\ref{Pbx}) on $x_i:=\lambda_{0,i}/\lambda_{i,0}$.

Note also that all $\lambda_{i,j}$ Poisson commute with the hole circumference $P$, and, since the twist coordinate $\tau_P$ constructed in \cite{Ch20-2} depends only on quotients $\lambda_B/\lambda_{0,0}$, all $\lambda_{i,j}$ Poisson commute with $\tau_P$ as well, so the semiclassical algebra of $\lambda_{i,j}$ completely splits from that of the rest of the Riemann surface.

\begin{figure}[tb]
\begin{pspicture}(-6,-4)(6,4){
\newcommand{\PATTERN}{%
{\psset{unit=1}
\rput{-45}(-3,-2.5){\psellipse[linecolor=blue, linewidth=1pt](0,0)(.75,.5)}
\rput{30}(2.5,-2){\psellipse[linecolor=blue, linewidth=1pt](0,0)(1,.5)}
\rput{-5}(0.29,0){\psellipse[linecolor=blue,  linewidth=1pt](0,0)(2.85,0.3)\psframe[linecolor=white, fillstyle=solid, fillcolor=white](-2.9,0)(2.9,.6)\psellipse[linecolor=blue, linestyle=dashed,  linewidth=1pt](0,0)(2.85,.3)}
\psbezier[linecolor=blue](-2.47,-3.03)(-1.47,-2.03)(1.134,-1.634)(1.634,-2.5)
\psbezier[linecolor=blue](-3.53,-1.97)(-2.56,-1)(-2,1)(-3.35,2.35)
\psbezier[linecolor=blue](3.35,-1.5)(2.85,-0.6)(3.19,1.2)(3.69,2)
\psbezier[linecolor=blue](-1.5,1.6)(-1.56,0.6)(-2.35,1.35)(-3.35,2.35)
\psbezier[linecolor=blue](1.5,1.3)(1.56,0.4)(3.19,1.2)(3.69,2)
\psbezier[linecolor=blue](1.5,1.3)(1.46,0.4)(-1.46,0.6)(-1.5,1.6)
\psbezier[linecolor=blue](-1.3,2.8)(-1.4,1.8)(-2.35,1.35)(-3.35,2.35)
\psbezier[linecolor=blue](2,2.6)(2.06,1.6)(3.19,1.2)(3.69,2)
\psbezier[linecolor=blue](0,2.8)(-0.2,1.9)(-1.2,1.9)(-1.3,2.8)
\psbezier[linecolor=blue](0,2.8)(0.4,1.6)(1.66,1.8)(2,2.6)
\pscircle[linewidth=1pt,linestyle=dashed](-3.15,2.15){0.3}
\pscircle[linewidth=1pt,linestyle=dashed](0,2.6){0.2}
\pscircle[linewidth=1pt,linestyle=dashed](-1.5,1.45){0.15}
\pscircle[linewidth=1pt,linestyle=dashed](-1.3,2.5){0.3}
\pscircle[linewidth=1pt,linestyle=dashed](1.5,1.1){0.2}
\pscircle[linewidth=1pt,linestyle=dashed](2,2.3){0.3}
\pscircle[linewidth=1pt,linestyle=dashed](3.55,1.86){0.2}
}
}
\rput(-4,0){\rput(0,0){\PATTERN}
\rput{0}(0,0){
\rput(0,0){
\psbezier[linecolor=black,linestyle=dashed](1.44,1.27)(1.47,0.7)(1.42,0.7)(1.50,-0.35)
\psbezier[linecolor=black,linestyle=dashed](3.59,2)(3,1)(2.85,1)(2.75,-0.3)
\psbezier[linecolor=black,linestyle=dashed](-2.3,0.33)(-2.4,1)(-2.4,1)(-3.35,2.35)
\psbezier[linecolor=black,linestyle=dashed](-1.6,1.6)(-1.4,0.5)(-1.4,0.5)(-1.35,-0.12)
\psbezier[linecolor=black,linestyle=dashed](-1.35,2.5)(-1.2,1.5)(-1.15,1)(-1.1,0.35)
\psbezier[linecolor=black,linestyle=dashed](-0.1,2.8)(0,2)(0,1.5)(0,0.35)
\psbezier[linecolor=black,linestyle=dashed](1.9,2.3)(1.8,1.3)(1.8,1.3)(1.75,0.1)
}
}
\rput(-1,0){
\rput(2.6,-0.45){\makebox(0,0)[tc]{\hbox{\tiny$\Delta_n$}}}
\rput(3.8,-0.4){\makebox(0,0)[tc]{\hbox{\tiny$\Delta_1$}}}
\rput(2.85,0.2){\makebox(0,0)[bl]{\hbox{\tiny$\Delta_2$}}}
\rput(1.2,0.35){\makebox(0,0)[bl]{\hbox{\tiny$\Delta_i$\large$\dots$}}}
\rput(0.1,0.35){\makebox(0,0)[bl]{\hbox{$\cdots$}}}
\rput(-1.25,0.4){\makebox(0,0)[bl]{\hbox{\tiny$\Delta_{n-2}$}}}
\rput(-0.5,-0.2){\makebox(0,0)[tc]{\hbox{\tiny$\Delta_{n-1}$}}}
}
\rput(9,0){{\PATTERN}
\rput{0}(0,0){\psbezier[linecolor=red](1.46,1.27)(1.44,0.4)(.75,-1.95)(1.15,-2.15)
\psbezier[linecolor=red](1.46,1.27)(1.48,0.4)(3,-0.4)(3.1,-1)
\psbezier[linecolor=red](1.15,-2.15)(1.55,-2.35)(3.3,-2.2)(3.1,-1)
\rput{-5}(0.29,0.3){\parametricplot[linecolor=magenta,linestyle=dashed,linewidth=1pt]{0}{180}{2.8 t cos mul  0.15 t sin  mul}
\psbezier[linecolor=magenta](1.1,1.05)(1.2,0.3)(2.75,-0.5)(2.8,0)
\psbezier[linecolor=magenta](1.1,1.05)(1.2,-0.2)(-2.95,-0.5)(-2.9,0)
}
\psbezier[linecolor=red](1.48,1.27)(1.52,0.4)(3.12,0.3)(3.22,0.8)
\psbezier[linecolor=red](1.48,1.27)(1.46,0.1)(-2.47,0.3)(-2.55,0.8)
\psbezier[linecolor=red](0,2.8)(0.3,1.2)(3.32,1.3)(3.22,0.8)
\psbezier[linecolor=red](0,2.8)(-0.3,0.6)(-2.63,1.3)(-2.55,0.8)
}
}
\rput(9,0){
\multiput(3.3,1.85)(-0.2,0.2){4}{\psline[linecolor=white,linewidth=4pt](0,0)(0.1,0.2)}
\multiput(-2.4,2.03)(0.2,0.035){6}{\psline[linecolor=white,linewidth=4pt](0,0)(-0.035,0.2)}
}
\rput(8,0){
\rput(4.3,0.6){\makebox(0,0)[lb]{\hbox{\tcr{\small$\lambda_{0,i}$}}}}
\rput(-1.6,0.6){\makebox(0,0)[rb]{\hbox{\tcr{\small$\lambda_{i,0}$}}}}
\rput(4.2,-0.1){\makebox(0,0)[lb]{\hbox{\textcolor{magenta}{\small$\lambda_{0,0}$}}}}
\rput(4.2,-1.1){\makebox(0,0)[lb]{\hbox{\textcolor{magenta}{\small$\lambda_{B}$}}}}
\rput(1.5,1.2){\makebox(0,0)[bl]{\hbox{$n{\sim} 0$}}}
\rput(1.2,2.7){\makebox(0,0)[bl]{\hbox{$i$}}}
}
}
}
\end{pspicture}
\caption{\small
A  ``crown'' of $m=7$ decorated bordered cusps attached to a pair of pants.}
\label{fi:n-bcsps}
\end{figure}

\subsection{The main results and the structure of paper}
In Sec.~\ref{s:action}, we propose a variant of the action on the ``fool's crown,'' (\ref{act1}), that seems to be natural by the following reasons: (i) it depends only on boundary structures: $\lambda$ lengths $\lambda_{i,i+1}$ of arcs $\mathfrak a_{i,i+1}$ between neighbor bordered cusps and lengths $s_i$ of horocycle segments confined between $\mathfrak a_{i-1,i}$ and $\mathfrak a_{i,i+1}$; (ii) it is the only such combination that is decoration-independent, so we refrain from introducing new characters into the play; (iii) upon choosing $\kappa=1$, the action functional $\int d\Omega\, e^{-S}$, where $d\Omega$ is the integration measure found in Lemma~\ref{lm:measure} that is compatible with the Poisson structure, becomes finite and analytic in the limit of coinciding bordered cusps, $\Delta_i\to \Delta_{i+1}$. Besides that, we also want this action to be at least quasi-additive, which means that adding new bordered cusps affects it only locally. Based on these considerations, we introduce in Theorem~\ref{th:volumes} the volumes $V^{\text{crown}}_{n,P}$ of moduli spaces of ``fool's crowns''---the main construction of this paper. We also derive the invariant symplectic form (\ref{omega}) for decoration-independent quantities $\xi_i$ previously introduced in \cite{Ch20-2}. 

We explore the continuum limit $n\to\infty$ of the corresponding action and measure in Section~\ref{s:Schwarzian}. We interpret the variables $\Delta_i$ as a discrete approximation of circle diffeomorphisms. We demonstrate that in this continuum limit, the action amended by the integration measure contains the ``standard'' Hill potential $-\frac12 S[f,t]+\frac12 [f'(t)]^2$, where $f'(t)$ describes a limiting density of distribution of $\Delta_i$ and $S[f,t]$ is the Schwarzian derivative, and the ``disc amplitude'' term $A_1^{(0)}\sim \int \log f'$, which can be attributed to the corresponding term in string B-models \cite{BKMP}. 

In Sec.~\ref{s:sympl}, we explore the continuum limit of the invariant symplectic form (\ref{omega}) showing that it contains both a standard Gelfand--Fuchs cocycle and an additional term, which combine into the symplectic form presented in Theorem~\ref{th:sympl}. The same form was derived from conformal mapping considerations by Alekseev and Meinrenken \cite{Al-Mein} and Stanford and Witten \cite{SW1}.

In Sec.~\ref{s:fat}, we rewrite the action in terms of shear coordinates on the fat graph represented by a circle with $n$ attached half-edges corresponding to bordered cusps. This action is given by a nice expression (\ref{action-shear}). It happen however that the invariant volume form $d\Omega$ becomes rather complicated (in contrast to the full invariant integration measure, which is just $\prod (dy_i d\alpha_i)$), so we do not proceed with evaluation of volumes $V^{\text{crown}}_{n,P}$ from the standpoint of shear coordinates noting however that these volumes are clearly finite. 

In Sec.~\ref{s:disc}, we derive volumes $V^{\text{disc}}_{n}$ of moduli spaces of a disc $\Sigma_{0,1,n}$ and consider their continuum limit. Upon fixing the global $SL_2(\mathbb R)$ symmetry, the effective variables $z_i$, $i=2,3,\dots, n-2$, now belong to the interval $[0,1]$ (we set $\{z_1,z_{n-1}, z_n\}=\{0,1,\infty\}$). The formula in Proposition~\ref{pr:volumes-disc} is rather similar to the one in the ``fool's crown'' case. If we do not exponentiate variables $z_i$, then the action in the continuum limit does not contain the term quadratic in $f'(t)$.

We present some development ideas in the conclusion. In Appendix~A, we collect formulas for fool's crown volumes in two asymptotic regimes, $P\to\infty$ and $P\to 0$. In Appendix~B, we present a standard evaluation of changing the Gelfand--Fuchs symplectic form (\ref{GF}) upon ``global'' variable changing.

\section{The action on a fool's crown}\label{s:action}
\setcounter{equation}{0}
\subsection{Deriving the action}
Let $\lambda_{i,j}$ be the lambda length (exponential of the signed half the geodesic length between intersections with decorating horocycles) of the arc $\mathfrak a_{i,j}$ between $i$th and $j$th border cusps counted counterclockwise. Then, e.g., $\lambda_{j,i}$ is the lambda-length of another arc $\mathfrak a_{j,i}$ joining the same cusps in opposite direction. We let $s_i$ denote the length of the horocycle arc confined between two points of intersections with the respective arcs $\mathfrak a_{i-1,i}$ and $\mathfrak a_{i,i+1}$. The combination invariant w.r.t. the decoration choice is then
\be\label{inv}
\prod_{i=1}^n \lambda_{i,i+1}s_i,
\ee
which is easy to see from the following: let the base of a horocycle be at infinity; the horocycle is then just a horizontal line parallel to the $x$-axis, and arcs terminating at infinity are vertical lines. Due to the global projective invariance, we can set the horocycle line to be $y=1$ with $s_i=z_2-z_1$, the (Euclidean) length of the horizontal interval confined between two arcs in the picture below. If we move this line to $y=e^a$, then $s_i\to s_i e^{-a}$ whereas each of the $\lambda$-lengths of the bounding arcs will be multiplied by $e^{a/2}$, so the above combination remains unchanged:
$$
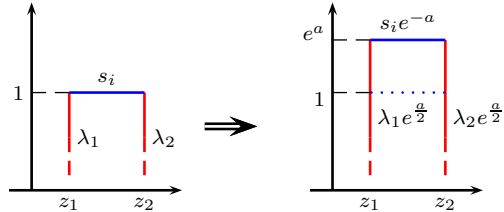
\begin{pspicture}(-4,-0.5)(4,2.5){\psset{unit=1}
\rput(-3,0){
\rput(0,0){\psline[linewidth=1pt,linecolor=red](0.5,0.7)(0.5,1.3)}
\rput(0,0){\psline[linewidth=1pt,linecolor=red](1.5,0.7)(1.5,1.3)}
\rput(0,0){\psline[linewidth=1pt,linecolor=red,linestyle=dashed](0.5,0.2)(0.5,0.7)}
\rput(0,0){\psline[linewidth=1pt,linecolor=red,linestyle=dashed](1.5,0.2)(1.5,0.7)}
\rput(0,0){\psline[linewidth=1pt, linecolor=blue](0.5,1.3)(1.5,1.3)}
\rput(0,0){\psline[linewidth=1pt]{->}(-0.3,0)(2,0)}
\rput(0,0){\psline[linewidth=1pt]{->}(0,0)(0,2.5)}
\rput(0,0){\psline[linewidth=0.5pt,linestyle=dashed](0,1.3)(0.5,1.3)}
\rput(-0.1,1.3){\makebox(0,0)[rc]{\tiny$1$}}
\rput(0.5,-0.1){\makebox(0,0)[tc]{\tiny$z_1$}}
\rput(1.5,-0.1){\makebox(0,0)[tc]{\tiny$z_2$}}
\rput(1,1.4){\makebox(0,0)[bc]{\tiny$s_i$}}
\rput(1.6,0.7){\makebox(0,0)[cl]{\tiny$\lambda_2$}}
\rput(0.6,0.7){\makebox(0,0)[cl]{\tiny$\lambda_1$}}
\psline[doubleline=true,linewidth=1pt, doublesep=1pt, linecolor=black]{->}(2.3,0.85)(3,0.85)
}
\rput(1,0){
\rput(0,0){\psline[linewidth=1pt,linecolor=red](0.5,0.7)(0.5,2)}
\rput(0,0){\psline[linewidth=1pt,linecolor=red](1.5,0.7)(1.5,2)}
\rput(0,0){\psline[linewidth=1pt,linecolor=red,linestyle=dashed](0.5,0.2)(0.5,0.7)}
\rput(0,0){\psline[linewidth=1pt,linecolor=red,linestyle=dashed](1.5,0.2)(1.5,0.7)}
\rput(0,0){\psline[linewidth=1pt, linecolor=blue](0.5,2)(1.5,2)}
\rput(0,0){\psline[linewidth=1pt, linecolor=blue,linestyle=dotted](0.5,1.3)(1.5,1.3)}
\rput(0,0){\psline[linewidth=1pt]{->}(-0.3,0)(2,0)}
\rput(0,0){\psline[linewidth=1pt]{->}(0,0)(0,2.5)}
\rput(0,0){\psline[linewidth=0.5pt,linestyle=dashed](0,1.3)(0.5,1.3)}
\rput(0,0){\psline[linewidth=0.5pt,linestyle=dashed](0,2)(0.5,2)}
\rput(-0.1,1.3){\makebox(0,0)[rt]{\tiny$1$}}
\rput(-0.1,2){\makebox(0,0)[rb]{\tiny$e^a$}}
\rput(0.5,-0.1){\makebox(0,0)[tc]{\tiny$z_1$}}
\rput(1.5,-0.1){\makebox(0,0)[tc]{\tiny$z_2$}}
\rput(1,2.1){\makebox(0,0)[bc]{\tiny$s_ie^{-a}$}}
\rput(1.6,1){\makebox(0,0)[cl]{\tiny$\lambda_2e^{\frac a2}$}}
\rput(0.6,1){\makebox(0,0)[cl]{\tiny$\lambda_1e^{\frac a2}$}}
}
}
\end{pspicture}
$$
 
Here and hereafter, if not explicitly stated opposite, all enumeration is done mod $n$. For the action to be (quasi)additive,\footnote{Quasi-additivity here means that when adding a new bordered cusp, the action gets only ``local'' addition.} we have to take the logarithm of expression (\ref{inv}), so
\be\label{act1}
S=\kappa \log\left[ \prod_{i=1}^n \lambda_{i,i+1}s_i \right],
\ee
where $\kappa$ is the coupling constant, which, as we show below, is determined unambiguously from the requirement of analyticity of volumes of moduli spaces.

For odd $n=2k+1$, we have a unique choice of $s_i$ such that all $\lambda_{i,i+1}=1$, which amounts to a ``gauge fixing" in physical terms. In this case, all horocycles are ``kissing circles."

For even $n=2k$, a configuration with all horocycles ``kissing'' generally does not exist (it exists if and only if $\sum_i \delta_{2i}=\sum_i \delta_{2i-1}$ with $\delta_j$ defined below), but the action is nevertheless well defined.

Let $\Delta_i$ be the distance (along the hole perimeter) between the orthogonal projection of the zeroth ($=n$th) cusp to the hole perimeter and the projection of the $i$th cusp. (In Fig.~\ref{fi:n-bcsps}, we use $\Delta_i$ also to denote the corresponding points on the hole perimeter.) We let
\be\label{delta}
\delta_i:=\Delta_i - \Delta_{i-1}> 0,\quad i=1,\dots,n
\ee
denote the positive distance between neighbor endpoints. Obviously, $\sum_{i=1}^n \delta_i=P$, where $P>0$ is the hole circumference. (We use the term ``perimeter'' for denoting the shortest closed geodesic homeomorphic to the hole boundary. The hole circumference is the length of this geodesics.) Our first goal is to express the action (\ref{act1}) in terms of $\delta_i$ alone. 

\begin{lemma}\label{lm:delta}
The action has the form
$$
S=\kappa \left( \sum_{i=1}^n \log(e^{\delta_{i+1}+\delta_i}-1) - \sum_{i=1}^n \log(e^{\delta_i}-1)\right).
$$
\end{lemma}
\noindent
{\bf proof} We use some well-known formulas from hyperbolic geometry. The first one is the geodesic length $\ell$ of the arc of a horocycle confined between the uppermost point and the point at the angle $\theta$:
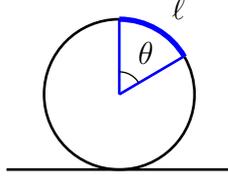
$$
\begin{pspicture}(-2,-0.5)(2,2){\psset{unit=1}
\rput(0,0){
%
%
\rput(0,0){\psline[linewidth=1pt](-1.5,0)(1.5,0)}
\rput(0,0){\psline[linewidth=1pt, linecolor=blue](0,1)(0,2)}
\rput(0,0){\psline[linewidth=1pt, linecolor=blue](0,1)(0.85,1.5)}
%
\rput(0,0){\psarc[linewidth=2pt,linecolor=blue](0,1){1}{30}{90}}
\rput(0,0){\psarc[linewidth=0.5pt](0,1){0.3}{30}{90}}
\rput(0,0){\psarc[linewidth=1pt](0,1){1}{90}{390}}
\rput(0.25,1.4){\makebox(0,0)[lb]{$\theta$}}
\rput(0.7,2){\makebox(0,0)[lb]{$\ell$}}
}
}
\end{pspicture}
$$
We have $\ell=\tan(\theta/2)$. 

{\bf Case 1.} We first consider the case of odd $n$. In the gauge where all $\lambda_{i,i+1}=1$ we have the pattern in the figure below: the hole perimeter is the vertical line ($y$-axis with identification $y\sim ye^P$), the bordered cusps are situated on the horizontal line ($x$-axis), and horocycles are kissing circles of radii $r_i$ based at the bordered cusps.
$$
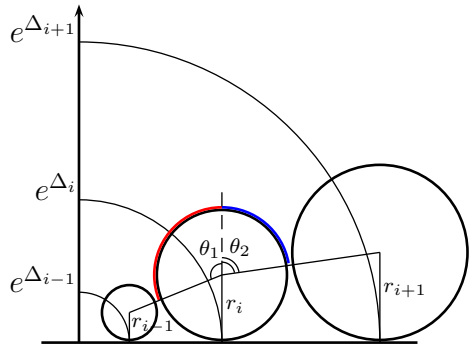
\begin{pspicture}(-1,0)(5,5){\psset{unit=1}
\rput(0,0){
%
%
\rput(0,0){\psline[linewidth=1pt](-0.5,0)(4.5,0)}
\rput(0,0){\psline[linewidth=1pt]{->}(0,0)(0,4.5)}
\rput(0,0){\pscircle[linewidth=1pt](0.67,0.4){0.4}}
\rput(0,0){\pscircle[linewidth=1pt](1.9,0.9){0.9}}
\rput(0,0){\pscircle[linewidth=1pt](4,1.2){1.2}}
\rput(0,0){\psarc[linewidth=0.5pt](0,0){0.67}{0}{90}}
\rput(0,0){\psarc[linewidth=0.5pt](0,0){1.9}{0}{90}}
\rput(0,0){\psarc[linewidth=0.5pt](0,0){4}{0}{90}}
\rput(0,0){\psarc[linewidth=1pt,linecolor=red](1.9,0.9){0.9}{90}{202}}
\rput(0,0){\psarc[linewidth=0.5pt](1.9,0.9){0.15}{90}{202}}
\rput(0,0){\psarc[linewidth=1pt,linecolor=blue](1.9,0.9){0.9}{10}{90}}
\rput(0,0){\psarc[linewidth=0.5pt](1.9,0.9){0.17}{10}{90}}
\rput(0,0){\psarc[linewidth=0.5pt](1.9,0.9){0.23}{10}{90}}
%
\rput(-0.02,0.67){\makebox(0,0)[rb]{$e^{\Delta_{i-1}}$}}
\rput(-0.02,1.9){\makebox(0,0)[rb]{$e^{\Delta_{i}}$}}
\rput(-0.02,4){\makebox(0,0)[rb]{$e^{\Delta_{i+1}}$}}
\rput(0,0){\psline[linewidth=0.5pt](0.67,0)(0.67,0.4)}
\rput(0,0){\psline[linewidth=0.5pt](1.9,0)(1.9,0.9)}
\rput(0,0){\psline[linewidth=0.5pt, linestyle=dashed](1.9,0.9)(1.9,2)}
\rput(0,0){\psline[linewidth=0.5pt](4,0)(4,1.2)}
\rput(0,0){\psline[linewidth=0.5pt](1.9,0.9)(0.67,0.4)}
\rput(0,0){\psline[linewidth=0.5pt](1.9,0.9)(4,1.2)}
\rput(1.88,1.13){\makebox(0,0)[rb]{\tiny$\theta_1$}}
\rput(2,1.15){\makebox(0,0)[lb]{\tiny$\theta_2$}}
\rput(0.7,0.13){\makebox(0,0)[lb]{\tiny$r_{i-1}$}}
\rput(1.95,0.4){\makebox(0,0)[lb]{\tiny$r_{i}$}}
\rput(4.05,0.6){\makebox(0,0)[lb]{\tiny$r_{i+1}$}}
}
}
\end{pspicture}
$$
Simple trigonometric formulas give
$$
e^{\Delta_{i+1}}-e^{\Delta_i}=2\sqrt{r_ir_{i+1}},\quad e^{\Delta_{i}}-e^{\Delta_{i-1}}=2\sqrt{r_{i-1}r_{i}}
$$
and
$$
\tan(\theta_1/2)=\frac{-\cos(\theta_1)+1}{\sin(\theta_1)}
$$
where $\cos(\theta_1)=\frac{r_{i-1}-r_{i}}{r_{i-1}+r_{i}}$ and $\sin(\theta_1)=\frac{e^{\Delta_{i}}-e^{\Delta_{i-1}}}{r_{i-1}+r_{i}}=\frac{2\sqrt{r_ir_{i-1}}}{r_{i-1}+r_{i}}$, so
$$
\tan(\theta_1/2)=\sqrt{r_i/r_{i-1}}.
$$
Analogously,
$$
\tan(\theta_2/2)=\sqrt{r_i/r_{i+1}}
$$
and therefore
\be
s_i=\tan(\theta_1/2)+\tan(\theta_2/2)=\sqrt{r_i/r_{i-1}}+\sqrt{r_i/r_{i+1}}=\frac{\sqrt{r_ir_{i-1}}+\sqrt{r_ir_{i+1}}}{\sqrt{r_{i-1}r_{i+1}}}=\frac{e^{\Delta_{i+1}}-e^{\Delta_{i-1}}}{2\sqrt{r_{i-1}r_{i+1}}}
\ee
Taking now the product, we obtain
$$
\prod_{i=1}^n s_i=\prod_{i=1}^n \bigl(e^{\Delta_{i+1}}-e^{\Delta_{i-1}}\bigr)/\left(2^n\prod_{i=1}^n r_i\right)=\frac{\prod_{i=1}^n \bigl(e^{\Delta_{i+1}}-e^{\Delta_{i-1}}\bigr)}{\prod_{i=1}^n \bigl(e^{\Delta_{i}}-e^{\Delta_{i-1}}\bigr)}=\prod_{i=1}^n \frac{e^{\delta_{i+1}+\delta_i}-1}{e^{\delta_i}-1},
$$
and taking the logarithm, we obtain the desired formula. Case 1 is proved.

{\bf Case 2.} We now consider the case of even $n$. We can still make $\lambda_{i,i+1}=1$ for arcs joining neighbor bordered cusps with $i=1,\dots,n-1$. For the last arc with $\lambda_{n,1}=e^{\ell/2}$ we have the following pattern:   
$$
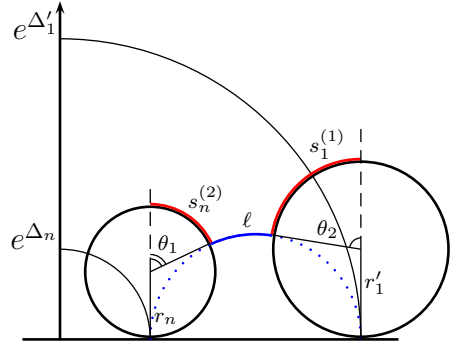
\begin{pspicture}(-1,0)(5,5){\psset{unit=1}
\rput(0,0){
%
%
\rput(0,0){\psline[linewidth=1pt](-0.5,0)(4.5,0)}
\rput(0,0){\psline[linewidth=1pt]{->}(0,0)(0,4.5)}
\rput(0,0){\pscircle[linewidth=1pt](1.2,0.9){0.9}}
\rput(0,0){\pscircle[linewidth=1pt](4,1.2){1.2}}
%
\rput(0,0){\psarc[linewidth=0.5pt](0,0){1.2}{0}{90}}
\rput(0,0){\psarc[linewidth=0.5pt](0,0){4}{0}{90}}
\rput(0,0){\psarc[linewidth=0.5pt](4,1.2){0.15}{90}{170}}
\rput(0,0){\psarc[linewidth=1pt,linecolor=red](1.2,0.9){0.9}{25}{90}}
\rput(0,0){\psarc[linewidth=1pt,linecolor=red](4,1.2){1.2}{90}{170.5}}
\rput(0,0){\psarc[linewidth=0.5pt](1.2,0.9){0.17}{25}{90}}
\rput(0,0){\psarc[linewidth=0.5pt](1.2,0.9){0.23}{25}{90}}
\rput(0,0){\psarc[linewidth=1pt,linecolor=blue, linestyle=dotted](2.6,0){1.4}{0}{180}}
\rput(0,0){\psarc[linewidth=1pt,linecolor=blue](2.6,0){1.4}{81}{115}}
%
\rput(-0.02,1.2){\makebox(0,0)[rb]{$e^{\Delta_{n}}$}}
\rput(-0.02,4){\makebox(0,0)[rb]{$e^{\Delta'_{1}}$}}
%
\rput(0,0){\psline[linewidth=0.5pt](2,1.28)(1.2,0.9)}
\rput(0,0){\psline[linewidth=0.5pt](1.2,0)(1.2,0.9)}
\rput(0,0){\psline[linewidth=0.5pt, linestyle=dashed](1.2,0.9)(1.2,2)}
\rput(0,0){\psline[linewidth=0.5pt](4,0)(4,1.2)}
\rput(0,0){\psline[linewidth=0.5pt](2.82,1.39)(4,1.2)}
\rput(0,0){\psline[linewidth=0.5pt, linestyle=dashed](4,1.2)(4,2.8)}
%
\rput(3.7,1.35){\makebox(0,0)[rb]{\tiny$\theta_2$}}
\rput(1.3,1.15){\makebox(0,0)[lb]{\tiny$\theta_1$}}
\rput(2.55,1.5){\makebox(0,0)[rb]{\tiny$\ell$}}
\rput(1.7,1.7){\makebox(0,0)[lb]{\tiny$s_n^{(2)}$}}
\rput(3.8,2.4){\makebox(0,0)[rb]{\tiny$s_1^{(1)}$}}
%
\rput(1.25,0.2){\makebox(0,0)[lb]{\tiny$r_{n}$}}
\rput(4.05,0.6){\makebox(0,0)[lb]{\tiny$r'_{1}$}}
}
}
\end{pspicture}
$$
where $\Delta'_1=\Delta_{n+1}=\Delta_1+P$ and $r'_i\equiv r_{n+i}=r_i\, e^P$. Circles with radii $r_n$ and $r'_1$ would be ``kissing'' if we enlarge one of radii $r_n$ or $r_1$ by $e^\ell$. We have therefore the set of relations:
$$
s_n^{(2)}=e^{-\ell/2}\sqrt{r_n/r'_1},\qquad s_1^{(1)}=e^{-\ell/2}\sqrt{r'_1/r_n}, \qquad e^{\Delta'_1}-e^{\Delta_n}=2\sqrt{r'_1r_ne^\ell}.
$$
The total action is then given by the sum
\begin{align*}
S&=\sum_{i=2}^{n-1}\log\left[\sqrt{\frac{r_i}{r_{i-1}}} + \sqrt{\frac{r_i}{r_{i+1}}}  \right] +\log \left[\sqrt{\frac{r_n}{r_{n-1}}} +e^{-\ell/2} \sqrt{\frac{r_n}{r'_{1}}}  \right] +\log \left[\sqrt{\frac{r'_1}{r'_{2}}} +e^{-\ell/2} \sqrt{\frac{r'_1}{r_{n}}}  \right] +\ell/2\\
&=\sum_{i=2}^{n-1}\log\frac{\sqrt{r_ir_{i+1}}+\sqrt{r_ir_{i-1}}}{\sqrt{r_{i+1}r_{i-1}}} + \log\frac{\sqrt{r_nr'_{1}}+e^{-\ell/2}\sqrt{r_nr_{n-1}}}{\sqrt{r_{n-1}r'_1}}+ \log\frac{\sqrt{r_nr'_{1}}+e^{-\ell/2}\sqrt{r'_1r'_{2}}}{\sqrt{r_{n}r'_2}}+\ell/2\\
&=\sum_{i=2}^{n-1}\log\bigl(e^{\Delta_{i+1}}-e^{\Delta_{i-1}} \bigr) +\log\bigl(e^{\Delta'_{1}}-e^{\Delta_{n-1}} \bigr)+ \log\bigl(e^{\Delta'_{2}}-e^{\Delta_{n}} \bigr)-\log\Bigl[ 2^n\,e^P\prod_{i=1}^n r_i\Bigr] -\ell/2\\
&=\sum_{i=1}^{n}\log\bigl(e^{\Delta_{i+2}}-e^{\Delta_{i}} \bigr) - \sum_{i=1}^{n}\log\bigl(e^{\Delta_{i+1}}-e^{\Delta_{i}} \bigr)
=\sum_{i=1}^{n}\log\bigl(e^{\delta_{i+2}+\delta_{i+1}}-1 \bigr) - \sum_{i=1}^{n}\log\bigl(e^{\delta_{i}}-1 \bigr),
\end{align*}
which completes the proof.

\subsection{Volumes of moduli spaces of ``fool's crowns''}

We consider first the case of odd $n$. We assume that volumes of moduli spaces of crowns are given by the standard formula
\be\label{volume}
V^{\text{crown}}_{n,P}=\int d\Omega\, e^{-S}
\ee
with $S$ the action and $d\Omega$ the Weil--Petersson volume form to be derived. 

In \cite{Ch20-2} a convenient set of decoration-independent coordinates on the fool's crown was introduced:
\be
x_i:=\lambda_{0,i}/\lambda_{i,0},\quad i=1,\dots,n-1.
\ee
The Poisson bracket derived from the Goldman bracket reads
\be\label{Pbx}
\{x_i,x_j\}=(x_j-x_i)x_i,\quad 1\le i<j\le n-1
\ee
Below we invert this bracket (it is nondegenerate for odd $n$) in order to obtain the invariant symplectic form. But for the purpose of deriving volumes, it suffices to evaluate the Pfaffian of the matrix
$$
A_{i,j}:=x_i(x_j-x_i),\ i<j;\quad A_{j,i}=-A_{i,j}
$$
\begin{lemma}\label{lm:measure}
$$
\hbox{Pf\,}A=x_1\prod_{i=2}^{n-1}(x_i-x_{i-1}),\quad (\hbox{we set\ } x_0:=0).
$$
The invariant measure is therefore
$$
d\Omega =\dfrac{ \prod_{i=1}^{n-1}dx_i }{x_1(x_2-x_1)\cdots (x_i-x_{i-1})\cdots (x_{n-1}-x_{n-2}) }.
$$
\end{lemma}
The {\bf proof} is merely the observation that substituting $x_i=x_{i-1}$ for $i>1$ results in that the $i$th and $(i-1)$th columns as well as the corresponding rows of the matrix $A$ become identical, so the Pfaffian is proportional to $x_{i}-x_{i-1}$. A mere substitution $x_1=0$ makes all entries in the first row and column equal zero.

It remains only to express the invariant measure in terms of $\Delta_i$ (or $\delta_i$). We show that
\be\label{x-Delta}
x_i=\frac{\sinh(\Delta_i/2)}{\sinh(P/2-\Delta_i/2)}
\ee
For this, turn again to geometry.
$$
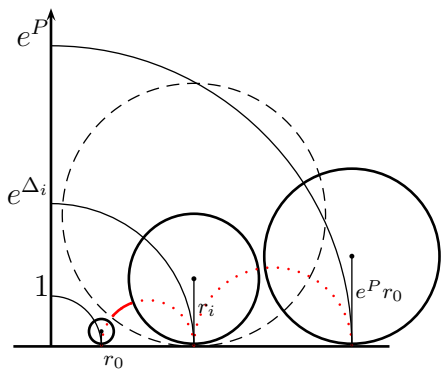
\begin{pspicture}(-1,-1)(5,5){\psset{unit=1}
\rput(0,0){
%
%
\rput(0,0){\psline[linewidth=1pt](-0.5,0)(4.5,0)}
\rput(0,0){\psline[linewidth=1pt]{->}(0,0)(0,4.5)}
\rput(0,0){\pscircle[linewidth=1pt](0.67,0.2){0.2}}
\rput(0,0){\pscircle[linewidth=0.5pt, fillstyle=solid, fillcolor=black](0.67,0.2){0.02}}
\rput(0,0){\pscircle[linewidth=1pt](1.9,0.9){0.9}}
\rput(0,0){\pscircle[linewidth=1pt, fillstyle=solid, fillcolor=black](1.9,0.9){0.02}}
\rput(0,0){\pscircle[linewidth=1pt](4,1.2){1.2}}
\rput(0,0){\pscircle[linewidth=1pt, fillstyle=solid, fillcolor=black](4,1.2){0.02}}
\rput(0,0){\pscircle[linewidth=0.5pt,linestyle=dashed](1.9,1.75){1.75}}
\rput(0,0){\psarc[linewidth=0.5pt](0,0){0.67}{0}{90}}
\rput(0,0){\psarc[linewidth=0.5pt](0,0){1.9}{0}{90}}
\rput(0,0){\psarc[linewidth=0.5pt](0,0){4}{0}{90}}
\rput(0,0){\psarc[linewidth=1pt,linecolor=red,linestyle=dotted](1.285,0){0.615}{0}{180}}
\rput(0,0){\psarc[linewidth=1pt,linecolor=red](1.285,0){0.615}{110}{140}}
\rput(0,0){\psarc[linewidth=1pt,linecolor=red,linestyle=dotted](2.95,0){1.05}{0}{180}}
%
\rput(-0.02,0.67){\makebox(0,0)[rb]{$1$}}
\rput(-0.02,1.9){\makebox(0,0)[rb]{$e^{\Delta_{i}}$}}
\rput(-0.02,4){\makebox(0,0)[rb]{$e^{P}$}}
\rput(0,0){\psline[linewidth=0.5pt](0.67,0)(0.67,0.2)}
\rput(0,0){\psline[linewidth=0.5pt](1.9,0)(1.9,0.9)}
\rput(0,0){\psline[linewidth=0.5pt](4,0)(4,1.2)}
%
%
\rput(0.7,-0.13){\makebox(0,0)[lt]{\tiny$r_{0}$}}
\rput(1.95,0.4){\makebox(0,0)[lb]{\tiny$r_{i}$}}
\rput(4.05,0.6){\makebox(0,0)[lb]{\tiny$e^Pr_{0}$}}
}
}
\end{pspicture}
$$
If we let $r'_i$ denote the radius of the horocycle kissing the horocycle based at $e^P$, then, obviously,
$$
x_i=\sqrt{r'_i/r_i},
$$
which implies
$$
x_i=\sqrt{\frac{r_0r'_i}{r_0r_i}}=e^{P/2}\frac{e^{\Delta_i}-1}{e^P-e^{\Delta_i}}=\frac{\sinh(\Delta_i/2)}{\sinh(P/2-\Delta_i/2)}.
$$
We then have that 
$$
x_i-x_{i-1} = \frac{\sinh(P/2)\sinh(\Delta_i-\Delta_{i-1})}{\sinh(P/2-\Delta_i/2)\sinh(P/2-\Delta_{i-1}/2)},
$$
and recalling that $\Delta_i-\Delta_{i-1}=\delta_i$, we obtain
$$
d\Omega=\prod_{i=1}^{n-1}\frac{d\Delta_i}{2} \left[\frac{\sinh(P/2)}{\sinh^2(P/2-\Delta_i/2)} \right]  \frac{\sinh^2(P/2-\Delta_1/2)\dots \sinh^2(P/2-\Delta_{n-2})\sinh(P/2-\Delta_{n-1})}{\sinh(\delta_1)\prod_{j=1}^{n-1}\bigl[ \sinh(P/2)\sinh(\delta_j/2) \bigr]},
$$
and therefore, since $P/2-\Delta_{n-1}/2=\delta_n/2$ and $\prod_{i=1}^{n-1}\frac{d\Delta_i}{2}=\prod_{i=1}^{n-1}\frac{d\delta_i}{2}$, we obtain
\be
d\Omega= \frac{\sinh(P/2)}{\prod_{i=1}^n \sinh(\delta_i/2)} \prod_{i=1}^{n-1}\frac{d\delta_i}{2}.
\ee
\begin{corollary}
In order for the volume $V^{\text{crown}}_{n,P}$ to be finite and limits $\Delta_i-\Delta_{i-1}\to 0^+$ to be analytic, we have to set
$$
\kappa=1.
$$
Then all factors $\sinh(\delta_i/2)$ in the measure are exactly canceled with the factors $e^{\delta_i}-1$ coming from the action.
\end{corollary}

\begin{theorem}\label{th:volumes}
The volumes $V^{\text{crown}}_{n,P}$ consistent with the Weil--Petersson symplectic form are given by the integrals
$$
V^{\text{crown}}_{n,P}=\frac{1}{2^{n-1}}\int_{\delta_1+\cdots +\delta_{n-1}\le P} \prod_{i=1}^{n-1}d\delta_i \frac{e^P-1}{\prod_{i=1}^n \bigl( e^{\delta_i+\delta_{i+1}}-1 \bigr)}
$$
with $\delta_i>0$, $\delta_{i+n}=\delta_i$, and $\delta_1+\cdots +\delta_{n-1}+\delta_n=P$. These volumes are finite for all $n$.
\end{theorem}

\begin{remark}
Volumes in question become finite for any $\kappa>0$. But only at the ``critical value'' $\kappa=1$, the nonanalyticity in the limit $\delta_i\to 0^+$ vanishes, so this value is clearly a ``chosen one.'' We therefore stick to this choice for the rest of this paper.
\end{remark}

\subsection{Mirzakhani volumes for $\mathcal M_{g,s,\mathbf n}$.}

Let $\mathcal M_{g,s,\mathbf n}$ be the moduli space of genus $g$ Riemann surfaces with $s>0$ holes enumerated from $1$ to $s$ and let $\mathbf n=(n_1,\dots,n_s)$ be the vector indicating the number of bordered cusps $n_i\in \mathbb Z_{+,0}$ attached to the $i$th hole. Then, since the only effect of adding bordered cusps to a boundary component on the modular group is the introduction of a new twist along the hole perimeter (which multiplies the volume by $P_i$---the circumference of the corresponding hole), volumes of the corresponding moduli spaces completely factorize and, denoting the original Mirzakhani volumes by $V_{g,s}^{\text{Mir}}(P_1,\dots,P_s)$, for the volumes of the moduli spaces $\mathcal M_{g,s,\mathbf n}$ we obtain the following simple lemma.
\begin{lemma}
The volumes $V_{g,s,\mathbf n}(P_1,\dots,P_s)$ of moduli spaces $\mathcal M_{g,s,\mathbf n}$ of Riemann surfaces of genus $g$ with $s$ holes with the fixed hole circumferences $P_1,\dots,P_s$ and with $n_i\ge 0$ bordered cusps attached to the boundary of the $i$th hole are given by the formula
$$
V_{g,s,\mathbf n}(P_1,\dots,P_s) = V_{g,s}^{\text{Mir}}(P_1,\dots,P_s) \prod_{i=1}^s f_{n_i}(P_i),\quad \hbox{where}\quad f_{n_i}(P_i)=\begin{cases} 1 & \hbox{for}\ n_i=0,\\ P_i V^{\text{crown}}_{n_i,P_i} & \hbox{for}\ n_i>0, \end{cases}
$$
where $V^{\text{crown}}_{1,P}=1$, $V^{\text{crown}}_{2,P}=\frac12\, \frac{P}{e^P-1}$, and for $n\ge 3$, the volumes $V^{\text{crown}}_{n,P}$ are given by Theorem~\ref{th:volumes}.  
\end{lemma}

\pagebreak

\subsection{Log-canonical variables for Poisson and symplectic relations}

\subsubsection{Variables $\xi_i$}

Following \cite{Ch20-2}, consider the combinations of $x_i$ (they were called $r_i$ in \cite{Ch20-2}, we change the notation here to distinguish them from the radii of horocycles)
\be\label{xi-i}
\xi_1:=x_1,\quad \xi_i:=x_i-x_{i-1},\ 2\le i\le n-1,
\ee
that is, take the same combinations of variables that had already appeared in the Pfaffian. 

We then have the lemma \cite{Ch20-2}
\begin{lemma}
The decoration-independent variables $\xi_i$ (\ref{xi-i}) have homogeneous Poisson relations
\be\label{xi-xi}
\{\xi_i,\xi_j\}=\xi_i\xi_j,\quad \hbox{or}\quad \{ \log(\xi_i),\log(\xi_j)\}=1\quad\hbox{for}\ \  1\le i<j\le n-1.
\ee
This algebra is nondegenerate for odd $n$ and has a unique Casimir element
\be\label{r-C}
C=\frac{\xi_1\xi_3\cdots \xi_{n-3} \xi_{n-1}}{\xi_2\xi_4\cdots \xi_{n-2}}
\ee
for even $n$.
\end{lemma}
Note that the variables $\xi_i$ take all values in $\mathbb R_+^{n-1}$, and for any given set of values of $\xi_i$ we have a unique configuration of cusps on the corresponding boundary component.

In the case of odd $n$, it is relatively easy to invert the skew-symmetric matrix $\mathcal A$ of size $(n-1)\times (n-1)$ with entries $\mathcal A_{i,j}=1$ for $i<j$. A standard exercise in linear algebra gives
\be
[\mathcal A^{-1}]_{i,j}=(-1)^{i+j}\quad \hbox{for}\ i<j;\qquad [\mathcal A^{-1}]_{i,j}=-[\mathcal A^{-1}]_{j,i}\ \forall \ i,j.
\ee
In the case of odd $n$, the invariant \emph{symplectic form} is therefore
\be\label{omega}
\omega = \sum_{1\le i<j\le n-1} (-1)^{i+j} d\log (\xi_i) \wedge d\log(\xi_j).
\ee
This symplectic form is equivalent to that introduced by Penner (\ref{WP-SS}).

Note that in the case of odd $n=2k+1$, the volume form $d\Omega$ in (\ref{volume}) coincides with that in the Duistermaat--Heckman theory:
\be\label{XXY}
d\Omega = \frac{\omega^k}{k!}.
\ee

\subsubsection{The case of even $n$}
Although Poisson relations for $\xi_i$ are degenerate for even $n$, and we cannot invert the Poisson matrix $\mathcal A$ (and therefore have no well-defined symplectic form), we still can define the \emph{volume form}. A similar situation takes place in the case of standard shear coordinates (logarithms of cluster variables $Z_\alpha$) for Riemann surfaces $\Sigma_{g,s}$ without bordered cusps: Poisson brackets of $\log Z_\alpha$ have Casimir elements, and no well-defined symplectic form exists, but the invariant volume form is just $d\Omega=\prod_\alpha d\log Z_\alpha$, where the product is taken over all $6g-6+3s$ variables corresponding to edges of a fat graph $\Gamma_{g,s}$, which is a spine of $\Sigma_{g,s}$ (see Sec.~\ref{s:fat} for an example of such fat graph for $\Sigma_{0,2,n}$); this volume form is explicitly invariant under flips (cluster mutations).

We therefore assume that 
\be\label{XXXY}
d\Omega =\prod_{i=1}^{n-1}\frac{d\xi_i}{\xi_i}
\ee
irrespectively on whether $n$ is odd or even. In the both cases, the corresponding volumes $V^{\text{crown}}_{n,P}$ are given by Theorem~\ref{th:volumes}.

\pagebreak

\subsection{Examples}
\subsubsection{ $V^{\text{crown}}_{3,P}$}
We present calculations of the first nontrivial example: the volume for a crown with three bordered cusps (the genuine ``fool's crown''). 

The integral to be evaluated is
$$
V_{3,P}=\frac14(e^P-1)\int_{0}^{P}d\Delta_2\int_0^{\Delta_2}d\Delta_1 \frac{1}{e^{\Delta_2}-1} \frac{1}{e^{P-\Delta_1}-1} \frac{1}{e^{P-\Delta_2+\Delta_1}-1}.
$$
Doing elementary integration over $\Delta_1$ and substituting $z=e^{\Delta_2}-1$, the remaining integral becomes
$$
\frac14 (e^P-1)\int_{0}^{e^P-1}\frac{dz}{z}\frac{1}{e^{2P}-1-z}\ln\left[ \frac{(z+1)(e^P-1)^2}{(e^P-1-z)^2}\right].
$$
Scaling now $z$ by $(e^P-1)$, we obtain the integral
$$
\int_0^1 \frac{dz}{z} \frac{1}{e^P+1-z}\ln\left[\frac{(e^P-1)z+1}{(1-z)^2} \right],
$$
which can be done using the dilogarithm function
$$
Li_2(x)=-\int_0^x d\xi\, \frac{\ln(1-\xi) }{\xi}=\sum_{n=1}^\infty \frac{1}{n^2} x^n.
$$
Using that $\frac{1}{z(e^{P}+1-z)}=\frac{1}{e^P+1}\Bigl(\frac 1z +\frac{1}{e^P+1-z} \Bigr)$, the result of integration can be written as a combination of dilogarithm functions,
\be\label{YY}
\frac14 \frac 1{e^P+1} \Bigl( -Li_2(1-e^P) +2Li_2(1+e^{-P}) -Li_2(1-e^{-2P}) +Li_2(1-e^{-P}) \Bigr).
\ee
We now simplify this expression using the celebrated five-term relation 
\be\label{five}
L(x)+L(y)-L(xy)-L\left[ \frac{x(1-y)}{1-xy}\right] -L\left[ \frac{y(1-x)}{1-xy}\right]=0
\ee
for Roger's function $L(x)$, which is by definition
$$
L(x):=\frac{\pi}{6}\left[ Li_2(x) +\frac 12\ln(x)\ln(1-x)\right].
$$
Choosing $x=1+e^{-P}$ and $y=1-e^{-P}$ and adding the proper logarithmic terms, we observe that all these terms are mutuallty cancelled, and the only terms that survive are
$$
\frac14\frac{1}{e^P+1}\left[ Li_2(1+e^{-P})+Li_2(1+e^P) \right]. 
$$
We now use the two-term relation, $Li_2(1-z)+Li_2(1-z^{-1})=\frac 12 \ln^2(z)$, which for negative $z=-e^{P}$ transforms into
$$
\frac 12 \ln(-e^P+i\epsilon)\ln(-e^P-i\epsilon)=\frac 12 (P+i\pi)(P-i\pi)=\frac 12 (P^2+\pi^2),
$$
and the final answer for the volume is therefore
\be\label{V3P}
V^{\text{crown}}_{3,P}= \frac{1}{4(e^P+1)} \frac{P^2+\pi^2}{2}.
\ee
We independently checked this answer in two limiting regimes, $P\to 0$ and $P\to\infty$ obtaining $\pi^2/16$ and $\frac 18 P^2e^{-P}$ in the respective cases (see Appendix~A). 

\begin{remark}
One should note a remarkable similarity between formula (\ref{V3P}) and the results for Mirzakhani volumes \cite{Mir06} of moduli spaces of hyperbolic Riemann surfaces with holes. Say, Mirzakhani's volume for a torus with one hole of perimeter $P$ is $\hbox{Vol}_{1,1}=\frac{\pi^2}{6}+\frac{P^2}{8}$, which, amending for volumes of discrete automorphism groups for two terms in the right-hand side ($2$ and $6$ in this case), finally gives $\hbox{Vol}_{1,1}=\frac{\pi^2}{12}+\frac{p^2}{48}$. However, in our case we have an additional exponential factor whose origin is not clear. 
\end{remark}

\subsubsection{$V^{\text{crown}}_{4,P}$} For even $n=2k$, the presence of the central element indicates the presence of flat directions: indeed, if we shift all odd $\Delta_{2i-1}$, $i=1,\dots, k$ by the same $\sigma$ w.r.t. all even $\Delta_{2i}$ (we assume, as above, that $\Delta_0=\Delta_{2k}=0$), then the value of the integrand in Theorem~\ref{th:volumes} remains unchanged, so we have a flat direction corresponding to the vector field $\sum_{i=1}^k \partial_{\Delta_{2i-1}}$. This does not however result in divergences since we have restrictions $\Delta_{2i-2}\le \Delta_{2i-1}\le \Delta_{2i}$.

The integral in question,
$$
\int_0^P \frac{d\Delta_2(e^P-1)}{(e^{\Delta_2}-1)(e^{P-\Delta_2}-1)}\int_0^{\Delta_2}d\Delta_1\int_{\Delta_2}^{P}d\Delta_3\frac{1}{(e^{\Delta_3-\Delta_1}-1)(e^{P+\Delta_1-\Delta_3}-1)},
$$
after integration over $\Delta_3$ becomes
$$
\int_0^P \frac{d\Delta_2\,e^{\Delta_2}}{(e^{\Delta_2}-1)(e^{P}-e^{\Delta_2})} \int_0^{\Delta_2}d\Delta_1 \log\left(\frac{(e^P-e^{\Delta_1})(e^{\Delta_1}-e^{\Delta_2-P})}{(e^{\Delta_2}-e^{\Delta_1})(e^{\Delta_1}-1)} \right),
$$
and integrating over $\Delta_1$, we obtain
$$
\int_0^P\!\! \frac{d\Delta_2\,e^{\Delta_2}}{(e^{\Delta_2}-1)(e^{P}-e^{\Delta_2})} \bigl[ -Li_2(e^{\Delta_2-P}) +Li_2(e^{-P}) -Li_2(e^{-\Delta_2}) -Li_2(e^{P}) +Li_2(e^{P-\Delta_2}) +Li_2(e^{\Delta_2})\bigr].
$$
This integral is obviously convergent, but we were unable to simplify it further.

\subsection{Other variants of the action}

We first have to mention the Schwarzian action proposed by Stanford and Witten in \cite{SW1}. Although the authors used discretization as a technical tool allowing implementation of some numerical methods to evaluate the action, the results (mostly collected in Appendix B) appeared to be rather close to those in this text. First, discretization of the integration measure ((2.16) in \cite{SW1}) $\prod_t \frac{df}{f'} $ has precisely the same form as the measure $d\Omega$ in Lemma~\ref{lm:measure}. Second, the authors proposed their variant of the Schwarzian discretization ((B.74) in \cite{SW1}), which also implements next-to-neighbor differences and uses the so-called conformal cross ratio:
$$
\frac{(f_{i+3}-f_{i+1})(f_{i+2}-f_i)}{(f_{i+3}-f_{i+2})(f_{i+1}-f_i)}\sim 4-2\varepsilon^2 S[f,t]+O(\varepsilon^3),
$$
where $S[f,t]$ is the Schwarzian derivative (\ref{Schw}). They also used the differences in the denominator of this expression to ensure the convergence of the integral (effectively erasing singularities of the integration measure), but, unlike in the above calculations, the limit $f_i\to f_{i+1}$ is wildely nonanalytic.

One more alternative approach to constructing an action on the ``fool's crowns'' is due to Alexander Goncharov and Zhe Sun \cite{Gon-Sun}. They propose to take $S=\beta\sum_i s_i$ with some (not fixed) positive $\beta$.  Whereas their approach has an advantage to be ``completely additive'' w.r.t. adding new bordered cusps by gluing new ideal triangles to the existing crown with the natural continuation of horocycles, their answer depends on decorations, and $\beta$ remains a free parameter. Note that volumes are finite in the both approaches, but in the approach advocated in this text, the volumes $V^{\text{crown}}_{n,P}$ are defined unambiguously by Theorem~\ref{th:volumes} and are decoration-independent, so we refrain from introducing extra degrees of freedom besides those appearing in discrete analogs of diffeomorphisms of a circle. 

\section{The action in the continuum limit}\label{s:Schwarzian}
\setcounter{equation}{0}

We now consider a continuum limit of integrals generating volumes $V^{\text{crown}}_{n,P}$. Such limits are, to some extent, a matter of art as no mathematically rigorous procedure of transition from discrete models to continuum integration exists (and it might be principally impossible to formulate one); however, in physical literature a functional integral is often described by discretization with subsequent transition to a theory with infinite number of degrees of freedom.

We explore only one among many possibilities of performing the continuum limit of integrals defining volumes $V^{\text{crown}}_{n,P}$: we assume that $\Delta_i$ are values of the function $f(t)$ with $t\in \mathbb R$ at the points $t_i=i/n$ and impose the quasiperiodicity condition  $f(t+1)=f(t)+P$. Then $f:[0,1]\to[0,P]$ describes the diffeomorphism $C^1\to C^1$ with $f(t)$ strictly increasing function with positive first derivative $f'(t)$.

The integration measure $\prod_i d\Delta_i$ then transforms into a standard (albeit ill-defined) measure of functional integration $\prod_t df(t)$.

Assuming smoothness of $f(t)$, we can proceed to a continuum limit of the action. Let $1/n=\varepsilon$ and $\delta_i=f(t_i)-f(t_i-\varepsilon)$. For the action, amended to the volume integration, we then obtain
\begin{align}
S&=\sum_{t_i}\log\bigl(e^{\delta_i+\delta_{i+1}}-1\bigr)\sim \frac{1}{\varepsilon} \int_0^1 dt\,\log\bigl(e^{f(t+\varepsilon)-f(t-\varepsilon)}-1\bigr)\nonumber\\
&= \frac{1}{\varepsilon}\int_0^1 dt\,\log\bigl(e^{2\varepsilon f'(t) +\frac16 2\varepsilon^3 f'''(t)+O(\varepsilon^5)}-1\bigr)\nonumber\\
&= \frac{1}{\varepsilon}\int_0^1 dt\,\log\Bigl( 2\varepsilon f'(t) + \frac12 (2\varepsilon)^2 [f'(t)]^2  +\frac16 2\varepsilon^3 f'''(t) + \frac16 (2\varepsilon)^3 [f'(t)]^3 + O(\varepsilon^4)\Bigr)\nonumber\\
&= \frac{\log(2\varepsilon)}{\varepsilon} + \frac{1}{\varepsilon}\int_0^1 dt\, \log (f'(t)) +\frac{1}{\varepsilon}\int_0^1 dt\,\log\Bigl[1+ \varepsilon f'(t) +\frac 16 \varepsilon^2 \frac{f'''(t)}{f'(t)} + \frac 46  \varepsilon^2 [f'(t)]^2 +O(\varepsilon^3)\Bigr]\nonumber\\
&=\frac{\log(2\varepsilon)}{\varepsilon} +\frac{1}{\varepsilon} \int_0^1 dt\, \log (f'(t)) +\int_0^1 dt\,\Bigl[ f'(t) - \frac {\varepsilon}{2}  [f'(t)]^2 +\frac 16 \varepsilon \frac{f'''(t)}{f'(t)} + \frac 46  \varepsilon [f'(t)]^2 +O(\varepsilon^2)  \Bigr]\nonumber\\
&=\frac{\log(2\varepsilon)}{\varepsilon}+ \frac{1}{\varepsilon}\int_0^1 dt\, \log (f'(t)) + P + \frac {\varepsilon}{6} \int_0^1 dt\,\Bigl[  \frac{f'''(t)}{f'(t)} + [f'(t)]^2\Bigr] +O(\varepsilon^2) 
\label{action}
\end{align}

\subsection{The Hill potential}
We are now identify all three nontrivial terms in the action formula (\ref{action}). Let us begin with two last terms with the factor $\frac {\varepsilon}{6}$. Both these terms had appeared in the formula for the Hill potential derived by Alekseev and Meinrenken (see Theorem~C in \cite{Al-Mein}). The same two terms had appeared in Stanford and Witten paper \cite{SW1}. The last term is just a standard quadratic term in the free-field action (and this term has a correct sign: recall that we integrate the exponent $e^{-S}$). The term with $f'''/f'$ is related to the Schwarzian derivative action
\be\label{Schw}
S[f,t]:=\frac{f'''_{ttt}}{f'_t}-\frac32 \left[\frac{f''_{tt}}{f'_t}\right]^2:
\ee
integrating $S[f,t]$ by parts, we obtain
$$
\int_0^1 dt\,S[f,t]= \int_0^1 dt\,\left[ \frac{f'''(t)}{f'(t)} +\frac32 f''(t)d\frac1 {f'(t)}\right]=\left.\frac 32 \,\frac{f''(t)}{f'(t)}\right|_0^1 -\frac12 \int_0^1 dt\,\frac{f'''(t)}{f'(t)}= -\frac12 \int_0^1 dt\,\frac{f'''(t)}{f'(t)},
$$
so the last two terms combine into
\be\label{Hill}
\frac {\varepsilon}{3}  \int_0^1 dt\,\Bigl[ -\frac12 S[f,t] +\frac 12 [f'(t)]^2\Bigr],
\ee
which, modulo opposite sign of $S[f,t]$, is the Hill potential (Theorem~C in \cite{Al-Mein}), which was given by the formula $\frac12 S[f,t] + \frac 12 c (f')^2$ with $c(t)$ related to the geodesic curvature of the bounding curve. For a possible origin of this sign mismatch, see the discussion in Sec.~\ref{s:sympl} below. Note however that in \cite{SW1}, the action in formula (1.1) for SYK model contains a combination $\frac{1}{2g^2}\bigl[ S[\phi,t] - (\phi')^2\bigr]$.

\subsection{The disc amplitude}
We now turn our attention to the first term, $\frac{1}{\varepsilon} \int dt\,\log(f'(t))$. Terms of this structure are commonly discarded, i.e., referred to the integration measure. For instance, this term will be cancelled if we consider a continuum limit of the whole action expression from Lemma~\ref{lm:delta}. But then the functional integration measure in turn becomes singular. We thus prefer to keep this term in the action in the continuum limit: we conjecturally identify it with the \emph{disc amplitude} introduced by Bouchard, Klemm, Marin\~o, and Pasquetti \cite{BKMP} in the context of the string B-model and mirror symmetry. For a spectral curve determined by the equation (usually non-alegbraic) $H(\rho,y)=0$, their disc amplitude reads \cite{BKMP}
\be\label{disc}
A_1^{(0)}=\int \log y \frac{d\rho}{\rho},
\ee
where the integral is taken over some specifically selected (closed or open) contour. If we take $\rho=e^t$ with identification $e^{t+1}\sim e^t$, then it seems natural to identify the variable $y$ with $f'(t)$ (this is a signature of \emph{topological recursion} models), and then the first term in action (\ref{action}) just describes the \emph{disc amplitude} of the trumpet upon ``erasing'' the hole at its end.

From the above standpoint, we may identify $\varepsilon$ parameter with an asymptotic expansion parameter $1/{N^2}=\varepsilon$, which is standard in random matrix models.

\begin{remark}
If we choose $y=\frac{df}{d\rho}$ instead, then the mismatch is just 
$$
\int \log\left[ \frac{d\rho}{dt}\right] \frac{d\rho}{\rho}=\int \log (\rho) \frac{d\rho}{\rho}=\frac 12 \int d\bigl[ \log\rho\bigr]^2\sim \frac 12,
$$
i.e., an irrelevant constant. So both possibilities produce virtually the same result.
\end{remark}

\begin{remark}
Note that the coefficient of the term $(f'(t))^2$ in the Hill potential is scheme-dependent: assume that we change $f(t)$ by $g(f(t))$ for some given function $g(f)$. Then from the transformation law of Schwarzian (\ref{Schw-tr}) (and it is not difficult to check it directly), for the action
$$
\sum_i \log (e^{g(f(t+\epsilon))-g(f(t-\epsilon))}-1)
$$
in the limit $\epsilon\to 0$, we obtain the expression
$$
\frac{1}{\epsilon}\int_0^1 \log[g'(f)f'(t)] dt + \frac{\epsilon}{6}\int_0^1 \left\{ [g'(f)]^2 [f'(t)]^2 - 2S[g,f] [f'(t)]^2 -2 S[f,t] \right\} dt
$$
So the coefficient of $[f']^2$ is given by $[g'(f)]^2 - 2 S[g,f]$. For example, if we just go from $e^f$ to $e^{i f}$, this term will change the sign. 
\end{remark}

\section{Fenchel--Nielsen coordinates in the continuum limit for the ``fool's crown"}\label{s:sympl}
\setcounter{equation}{0}
In this section, we explore the continuum limit of the symplectic form (\ref{omega}). The accounting in this section is based only on this form and on the relations (\ref{x-Delta}) and (\ref{xi-i}) between different sets of coordinates, whereas a choice of an action does not play any role. 

\subsection{The continuum limit of (\ref{omega})}

We first re-arrange terms in the symplectic form (\ref{omega}) Recall that $n=2q+1$ must be odd for this form to be defined, but in this calculation we also unfreeze the last variable $\xi_{n}=x_n-x_{n-1}$:
\begin{align*}
\omega &= \sum_{j=1}^{n-1}\left( \sum_{i=1}^{j-1} (-1)^{i+1} d\log (\xi_i)\right) \wedge d\log(\xi_j)(-1)^{j+1}\\
&= \sum_{k} \left[ d\log\Bigl( \frac{\xi_1}{\xi_2}\cdot \frac{\xi_3}{\xi_4}\cdots \frac{\xi_{2k-1}}{\xi_{2k}}\Bigr)\wedge d\log(\xi_{2k+1}) -
d\log\Bigl( \frac{\xi_1}{\xi_2}\cdot \frac{\xi_3}{\xi_4}\cdots \frac{\xi_{2k-1}}{\xi_{2k}}\cdot \xi_{2k+1}\Bigr)\wedge d\log(\xi_{2k+2}) \right]\\
&= \sum_{k} d\log\Bigl( \frac{\xi_1}{\xi_2}\cdot \frac{\xi_3}{\xi_4}\cdots \frac{\xi_{2k-1}}{\xi_{2k}}\Bigr)\wedge \bigl(d\log(\xi_{2k+1} -d\log(\xi_{2k}\bigr)\\
&= \sum_{k=1}^{q} d\log\Bigl( \frac{\xi_1}{\xi_2}\cdot \frac{\xi_3}{\xi_4}\cdots \frac{\xi_{2k-1}}{\xi_{2k}}\Bigr)\wedge d\log\Bigl(\frac {\xi_{2k+1}}{\xi_{2k}}\Bigr).
\end{align*}
We then assume that $\xi_j=g(t_j)-g(t_{j-1})>0$, where $g(t)$ describe a smooth mapping $C^1\to \mathbb R$ with $t_i=i/n$. We let $\varepsilon=1/n$ to be a small parameter. Under the assumption that $g'(t)>0$ for all $t$, expanding $\log(\xi_{2i-1}/\xi_{2i})$ around the point $t=t_{2i-1}$, we obtain
\begin{align}
\log (\xi_{2i-1}/\xi_{2i})\Bigm|_{t=t_{2i-1}}& =\log\frac{g(t)-g(t-\varepsilon)}{g(t+\varepsilon)-g(t)} = \log\left[\frac{g'(t)\varepsilon-\frac 12 g''(t)\varepsilon^2+O(\varepsilon^3)}{g'(t)\varepsilon+\frac 12 g''(t)\varepsilon^2+O(\varepsilon^3)}\right]\nonumber\\
&=\log\Bigl[ 1- \frac{g''(t)}{g'(t)}\varepsilon+O(\varepsilon^2)\Bigr]=- \frac{g''(t)}{g'(t)}\varepsilon+O(\varepsilon^2).\label{XX}
\end{align}
Analogously, when expanding around the point $t=2k$,
\be\label{XXX}
\log (\xi_{2k+1}/\xi_{2k})= \frac{g''(t_{2k})}{g'(t_{2k})}\varepsilon+O(\varepsilon^2)=\frac{d}{dt}\log\bigl(g'(t)\bigr)\Big|_{t=t_{2k}}\varepsilon+O(\varepsilon^2)
\ee
We now interpret the sum over $t_{2i-1}$ as a Riemann sum of the function $g''(t)/g'(t)$ with the sample points located at the midpoints of the intervals $[t_{2i-2},t_{2i}]$ (recall that $t_{2i}-t_{2i-2}=2\varepsilon$). Then we obtain that
\be\label{XXXX}
-\sum_{i=1}^{k} \frac{g''(t_{2i-1})}{g'(t_{2i-1})}\varepsilon = -\frac 12 \int_0^{t_{2k}} \frac{g''(t)}{g'(t)} dt + O(\varepsilon^2)=-\frac 12 \log\bigl(g'(t_{2k})\bigr) + \frac12 \log\bigl(g'(0)\bigr) + O(\varepsilon^2)
\ee
and combining (\ref{XXX}) and (\ref{XXXX}) and again interpreting the sum over $t_{2k}$ as a Riemann sum (which brings one extra factor of $1/2$), for the continuum limit $\varepsilon\to 0$ of the symplectic form (\ref{omega}) we therefore obtain the standard expression that follows from the Gelfand--Fuchs cocycle \cite{Al-Shat}, \cite{Wit1}:
\be\label{GF}
\omega=-\frac 14 \int_{0}^{1} d\bigl[ \log\bigl( g'(t)\bigr)  \bigr] \wedge d \left( \partial_t \bigl[ \log\bigl( g'(t)\bigr)  \bigr] \right) dt + O(\varepsilon).
\ee
Note that in this expression, the external differential $d$ acts in the infinite-dimensional configuration space of diffeomorphisms and, in particular, it commutes with the action of the spatial derivative $\partial_t$.

\subsection{Symplectic structure for the ``fool's crown''}

It is well-known how the form (\ref{GF}) transforms under functional variable changes; We nevertheless present this derivation in Appendix~B  for the reader's convenience. Our goal is to express this form in terms of the function $f(t_i):=\Delta_i$ for a function (\ref{x-Delta})
$$
g(f(t_i))=x_i=\frac{\sinh(f(t_i)/2)}{\sinh(P/2-\Delta_i/2)}=\frac{e^{\Delta_i}-1}{e^{P/2}-e^{\Delta_i-P/2}},
$$
so, we need to find $S[g,f]$ for $g=\frac{e^f-1}{e^{P/2}-e^{f-P/2}}$. It is well known that for any nondegenerate fractional-linear transformation and exponential $e^{\alpha f}$, we have that this Schwarzian is just a constant,
\be\label{SS}
S\left[ \frac{ae^{\alpha f}+b}{ce^{\alpha f}+d},\, f\right]=-\frac{\alpha^2}{2}\quad \forall a,b,c,d,\alpha \in \mathbb C.
\ee
This immediately follows from the composition law for $S[g,f]$: for $g(f(t))$, 
\be\label{Schw-tr}
S[g,t]=(f'_t)^2 S[g,f] + S[f,t];
\ee 
easy calculation then gives that $S[e^{\alpha f}, f]=-\alpha^2/2$, and possibly the most famous property of the Schwarzian derivative is that it is invariant under linear-fractional transformations, so it remains constant in this case.

For the continuous limit of the symplectic form $\omega$ in the ``fool's crown'' case, from (\ref{g-f}) we therefore obtain a theorem.
\begin{theorem}\label{th:sympl}
The Fenchel--Nielsen symplectic form of the ``fool's crown'' transforms in the continuum limit into the symplectic form
\begin{align}\label{omega-cont}
\omega&= -\frac14 \int_0^1 dt\, \Bigl[ d\bigl( \log f'(t)\bigr)\wedge \partial_t\, d \bigl( \log f'(t)\bigr)- df'(t) \wedge df(t)\Bigr]\nonumber\\
&=-\frac14 \int_0^1 dt\, \Bigl[ \frac{df'(t)\wedge df''(t)}{(f'(t))^2} - df'(t) \wedge df(t)\Bigr],
\end{align}
which is an extension of the standard 2-form derived from the Gelfand--Fuchs cocycle.
\end{theorem}

\subsection{Comparison with existing results}
The form (\ref{omega-cont}) exactly coincides (upon setting $\ell^2=1$) with the symplectic form derived by Alekseev and Meinrenken (Theorem~D in \cite{Al-Mein}) using the co-frame formalism. A similar expression had appeared in \cite{SW1} (formula (2.10) derived on the base of symplecticity of the coadjoint vector orbits under the action of Virasoro group of circle diffeomorphisms). However, the mutual sign between two terms in the symplectic form in \cite{SW1} is opposite as regarding these signs in \cite{Al-Mein} and in (\ref{omega-cont}). The reason for this mismatch is purely technical: the authors of \cite{SW1} used, instead of hyperbolic functions, the standard trigonometric functions $\sin$ and $\cos$; in this case, $\alpha^2=(i)^2=-1$ in (\ref{SS}) in contrast to $\alpha^2=1$ in our case, which results in inverting the sign of the second term. We suspect a similar, purely technical, reason for the mismatch between the variant of continuum action derived in Sec.~\ref{s:Schwarzian} (\ref{action}) and in Theorem~C of \cite{Al-Mein}.

\section{Fool's crowns and shear coordinates}\label{s:fat}
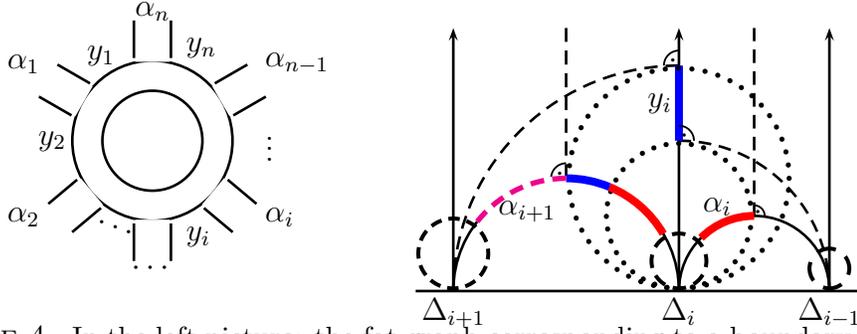
\begin{figure}[tb]
\begin{pspicture}(-6,-2)(6,2){
\newcommand{\ANGLE}[2]{
\psset{unit=#2}{\rput{#1}(0,0){\psarc[linewidth=0.7pt](0,0){0.2}{0}{90}
\rput(0.07,0.07){\makebox(0,0)[cc]{$\cdot$}}
}
}}
\rput(-4,0){
\rput(0,0){\pscircle[linewidth=1pt](0,0){0.7}}
\rput(0,0){\pscircle[linewidth=1pt](0,0){1.1}}
\rput{90}(0,0){\psline[linewidth=15pt](1.1,0)(1.6,0)
\psline[linewidth=13pt,linecolor=white](1.05,0)(1.65,0)}
\rput{30}(0,0){\psline[linewidth=15pt](1.1,0)(1.6,0)
\psline[linewidth=13pt,linecolor=white](1.05,0)(1.65,0)}
\rput{-40}(0,0){\psline[linewidth=15pt](1.1,0)(1.6,0)
\psline[linewidth=13pt,linecolor=white](1.05,0)(1.65,0)}
\rput{-90}(0,0){\psline[linewidth=15pt](1.1,0)(1.6,0)
\psline[linewidth=13pt,linecolor=white](1.05,0)(1.65,0)}
\rput{150}(0,0){\psline[linewidth=15pt](1.1,0)(1.6,0)
\psline[linewidth=13pt,linecolor=white](1.05,0)(1.65,0)}
\rput{220}(0,0){\psline[linewidth=15pt](1.1,0)(1.6,0)
\psline[linewidth=13pt,linecolor=white](1.05,0)(1.65,0)}
\rput(0,1.6){\makebox(0,0)[cb]{$\alpha_n$}}
\rput(-0.5,1){\makebox(0,0)[rb]{$y_1$}}
\rput(-1.5,0.9){\makebox(0,0)[rb]{$\alpha_1$}}
\rput(-1.15,0){\makebox(0,0)[rc]{$y_2$}}
\rput(-1.5,-0.9){\makebox(0,0)[rt]{$\alpha_2$}}
\rput{-22}(0,0){\rput(0,-1.35){\makebox(0,0)[cb]{$\cdots$}}}
\rput(0,-1.6){\makebox(0,0)[ct]{$\cdots$}}
\rput(0.45,-1.15){\makebox(0,0)[lt]{$y_i$}}
\rput(1.5,-0.9){\makebox(0,0)[lt]{$\alpha_i$}}
\rput(1.5,0){\makebox(0,0)[lc]{$\vdots$}}
\rput(1.5,0.9){\makebox(0,0)[lb]{$\alpha_{n-1}$}}
\rput(0.45,1.1){\makebox(0,0)[lb]{$y_{n}$}}
}
\rput(0,-2){
%
%
\rput(0,0){\psline[linewidth=1pt](-0.5,0)(5.5,0)}
\rput(0,0){\psline[linewidth=1pt]{->}(0,0)(0,3.5)}
\rput(0,0){\psline[linewidth=1pt]{->}(3,0)(3,3.5)}
\rput(0,0){\psline[linewidth=1pt]{->}(5,0)(5,3.5)}
\rput(0,0){\psarc[linewidth=1pt](1.5,0){1.5}{0}{180}}
\rput(0,0){\psarc[linewidth=1pt](4,0){1}{0}{180}}
\rput(0,0){\psline[linewidth=1pt,linestyle=dashed](1.5,1.5)(1.5,3.5)}
\rput(0,0){\psline[linewidth=1pt,linestyle=dashed](4,1)(4,3.5)}
\rput(0,0){\psarc[linewidth=1pt,linestyle=dashed](3,0){3}{90}{180}}
\rput(0,0){\psarc[linewidth=1pt,linestyle=dashed](3,0){2}{0}{90}}
\rput(0,0){\psline[linewidth=3pt,linecolor=blue](3,2)(3,3)}
\rput(0,0){\pscircle[linewidth=2pt,linestyle=dotted](3,1){1}}
\rput(0,0){\pscircle[linewidth=2pt,linestyle=dotted](3,1.5){1.5}}
\rput(0,0){\pscircle[linewidth=1.5pt,linestyle=dashed](3,0.4){0.4}}
\rput(0,0){\pscircle[linewidth=1.5pt,linestyle=dashed](0,0.5){0.5}}
\rput(0,0){\pscircle[linewidth=1.5pt,linestyle=dashed](5,0.3){0.3}}
\rput(3,2){\ANGLE{0}{1}}
\rput(3,3){\ANGLE{90}{1}}
\rput(1.5,1.5){\ANGLE{90}{1}}
\rput(4,1){\ANGLE{0}{0.7}}
%
\rput(0,0){\psarc[linewidth=3pt,linecolor=red](1.5,0){1.5}{30.5}{67.5}}
\rput(0,0){\psarc[linewidth=3pt,linecolor=blue](1.5,0){1.5}{67.5}{90}}
\rput(0,0){\psarc[linewidth=3pt,linecolor=red](4,0){1}{90}{135}}
%
%
\rput(0,0){\psarc[linewidth=2pt,linecolor=white](1.5,0){1.5}{90}{142}}
\rput(0,0){\psarc[linewidth=2pt,linecolor=magenta,linestyle=dashed](1.5,0){1.5}{90}{142}}
%
\rput(0.6,1.2){\makebox(0,0)[lt]{$\alpha_{i+1}$}}
\rput(3.7,1){\makebox(0,0)[rb]{$\alpha_{i}$}}
\rput(2.9,2.5){\makebox(0,0)[rc]{$y_{i}$}}
\rput(0,-0.1){\makebox(0,0)[ct]{$\Delta_{i+1}$}}
\rput(3,-0.1){\makebox(0,0)[ct]{$\Delta_{i}$}}
\rput(5,-0.1){\makebox(0,0)[ct]{$\Delta_{i-1}$}}
}
}
\end{pspicture}
\caption{\small In the left picture: the fat graph corresponding to a boundary component with $n$ boundary cusps: $\alpha$-variables are extended shear coordinates (defined in the right picture as a geodesic distances between perpendiculars to the bounding arcs and the horocycles lying to the right of the perpendicular if looking from inside the surface); $y_i$ are standard shear coordinates. In the right picture we show the pattern near the boundary cusps: vertical lines are infinite geodesics that all wind in the same direction to the hole perimeter; all these lines are approaching the same point on the absolute that we set to be $\infty$. Standard shear coordinates $y_i$ are geodesic distances between perpendiculars to the vertical lines separating the neighbor ideal triangles with vertices $\{\Delta_{i-1}, \Delta_i,\infty\}$ and $\{\Delta_{i+1}, \Delta_i,\infty\}$. The distances from the perpendiculars to the bounding arcs to the $i$th horocycle from the right and from the left are $\alpha_i$ and $\alpha_i+y_i$ respectively (intervals of the same length are painted the same color in the figure); the total length of the horocycle arc confined between $\mathfrak a_{i-1,i}$ and $\mathfrak a_{i,i+1}$ is therefore $e^{-\alpha_i}+e^{-\alpha_i+y_i}$.}
\label{fi:circle}
\end{figure}

\setcounter{equation}{0}
In this section, we are about to reformulate the approach explored in preceding sections in terms of (extended) shear coordinates of \cite{ChMaz2}. The fat graph $\Gamma_{0,2,n}$ we consider is just a circle with $n$ half-edges attached to the rim (from one side), see Fig.~\ref{fi:circle}. We have $n$ ``standard'' decoration-independent shear coordinates $y_i$, $i=1,\dots, n$, and $n$ ``extended'' decoration-dependent shear coordinates $\alpha_i$, $i=1,\dots, n$, one such coordinate per each bordered cusp. 

\subsection{The action}
For $\lambda_{i,i+1}$ we merely have $\lambda_{i,i+1} = e^{\alpha_i/2+\alpha_{i+1}/2+y_i/2}$, whereas for $s_i$ we have (see the right picture in  Fig.~\ref{fi:circle})
$$
s_i=e^{-\alpha_i}+e^{-\alpha_i+y_i},
$$
so the action $S$  (\ref{act1}) becomes
\be\label{action-shear}
S=\kappa\log\prod_{i=1}^n \Bigl[e^{-\alpha_i}(1+e^{-y_i}) e^{\alpha_i/2+\alpha_{i+1}/2+y_i/2}\Bigr] =  \kappa\log\prod_{i=1}^n \bigl(e^{y_i/2}+e^{-y_i/2} \bigr),\quad \sum_{i=1}^n y_i = \pm P.
\ee

\subsection{The integration measure}
It happens that the general formula for the integration measure consistent with the Poisson bracket is rather cumbersome. For $\lambda_{i,j}$ we have
$$
\lambda_{i,j} =e^{y_{i+1}/2+y_{i+2]/2+\cdots +y_j/2}}\bigl(1+e^{-y_j}+e^{-y_j-y_{j-1}}+\cdots + e^{-y_j-y_{j-1}-\cdots -y_{i+2}}\bigr),
$$
and for $\xi_i$ we then obtain after some algebra
\be
\xi_i=\frac{e^{P/2-y_2/2-\dots-y_n/2}\mu_1}{\mu_i\mu_{i+1}e^{y_{i+1}/2}},
\ee
where
\be
\mu_i:=e^{-(y_{i+1}+\cdots +y_n)/2}\bigl(1+e^{y_{i+1}}+e^{y_{i+1}+y_{i+2}}+\cdots + e^{y_{i+1}+\cdots +y_n}\bigr),\quad i=1,\dots, n-1.
\ee
In this expression we have already expressed $y_1=P-y_2-\cdots -y_n$, so the remaining $y_j$ are unrestricted and assume all values between $-\infty$ and $+\infty$. The resulting expression for $d\Omega$ remains rather complicated; we consider just one limiting case, $y_j\to +\infty$ for all $j=2,\dots, n$. Then 
$$
\lim_{y_k\to +\infty} \frac{d\xi_i}{\xi_i}=-\sum_{k=i+1}^{n} dy_k,
$$
so
$$
\mathop{\wedge}\limits_{i=1}^{n-1}\frac{d\xi_i}{\xi_i}\Bigm|_{y_k\to +\infty}=(-1)^{n-1} \mathop{\wedge}\limits_{i=2}^{n} dy_i,
$$
which clearly indicate that we have flattening directions of this measure. We can however prove that 
\be
|d\Omega|\le C \prod |dy_i|
\ee
in any direction, so the combination $d\Omega\, e^{-S}$ with action (\ref{action-shear}) is integrable for any $\kappa>0$. We nevertheless expect that the value $\kappa=1$ remains ``special'' in this case as well.

\section{Volumes of moduli spaces for the disc $\Sigma_{0,1,n}$}\label{s:disc}
\setcounter{equation}{0}

We now consider the case of a disc $\Sigma_{0,1,n}$ with $n\ge3$ boundary cusps, which we enumerate from $n=0$ to $n-1$ moving counterclockwise. The cusps labeled ``$0$'', ``$1$'', and ``$n-1$'' play a special role in the construction below. We now have unique arcs $\mathfrak a_{i,j}$ joining cusps with labels $i$ and $j$, and since arcs are not directed, we can always assume that $i<j$. Then, for (still not completely decoration-independent) combinations
\be
x_j:=\lambda_{1,j}/\lambda_{0,j},\quad j=2,\dots,n-1.
\ee
we obtain the Poisson bracket \cite{Ch20-2}
\be
\{x_i,x_j\}=x_ix_j-x_i^2,\quad 2\le i<j\le n-1,
\ee
so these relations coincide with brackets (\ref{Pbx}) of $x_i$ in a ``fool's crown.'' Then, taking
\be
\xi_i=\{x_2,\ i=2;\ x_i-x_{i-1}, \ 3\le i\le n-1\},
\ee
we obtain homogeneous Poisson relations $\{\xi_i,\xi_j\}=\xi_i\xi_j$ for $2\le i<j\le n-1$ and, finally, we can introduce completely decoration-independent variables
\be\label{r-disc}
\widehat \xi_i:=\xi_i/\xi_2,\quad 3\le i\le n-1,
\ee
for which we have the proposition \cite{Ch20-2}
\begin{proposition}
in the disc with $n\ge 4$ bordered cusps we have $n-3$  Fenchel--Nielsen coordinates $\widehat \xi_i$ (\ref{r-disc}) that are
decoration independent and have homogeneous Poisson relations
\be\label{r-r1}
\{\widehat \xi_i,\widehat \xi_j\}=\widehat \xi_i\widehat \xi_j,\quad 3\le i<j\le n-1;
\ee
this algebra is nondegenerate for odd $n$ and it has a unique Casimir element
\be\label{r-C1}
C=\frac{\widehat \xi_3\cdots \widehat \xi_{n-3}\widehat  \xi_{n-1}}{\widehat \xi_4\cdots\widehat  \xi_{n-2}}
\ee
for even $n$.
\end{proposition}

\subsection{The measure and the action}

We now proceed to evaluating the measure of integration compatible with the Poisson structure; for simplicity we consider only the case of odd $n$, but the result will stay the same for even $n$ as well.

Due to the global $SL(2,\mathbb R)$ symmetry, we have to fix positions of three boundary cusps on the (closure of the) real line: we choose
\be
z_{0}(=z_n)=\infty,\quad z_1=0,\quad z_{n-1}=1,\quad \hbox{and}\quad 0<z_i<z_j<1 \ \hbox{for}\ 2\le i<j \le n-2.
\ee
The pattern of kissing horocycles now takes the form
$$
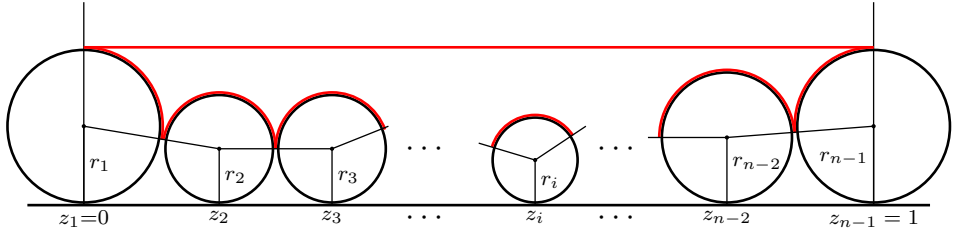
\begin{pspicture}(-1,-0.5)(10,3){\psset{unit=1.5}
\rput(0,0){
%
%
\rput(0,0){\psline[linewidth=1pt](-0.5,0)(7.5,0)}
\rput(0,0){\psline[linewidth=0.5pt](0,0)(0,1.8)}
\rput(0,0){\psline[linewidth=0.5pt](7,0)(7,1.8)}
\rput(0,0){\pscircle[linewidth=1pt](0,0.7){0.7}}
\rput(0,0){\pscircle[linewidth=1pt,fillstyle=solid, fillcolor=black](0,0.7){0.03}}
\rput(0,0){\pscircle[linewidth=1pt](1.2,0.5){0.5}}
\rput(0,0){\pscircle[linewidth=1pt,fillstyle=solid, fillcolor=black](1.2,0.5){0.03}}
\rput(0,0){\pscircle[linewidth=1pt](2.2,0.5){0.5}}
\rput(0,0){\pscircle[linewidth=1pt,fillstyle=solid, fillcolor=black](2.2,0.5){0.03}}
\rput(2.85,0.5){\makebox(0,0)[lc]{$\cdots$}}
\rput(0,0){\pscircle[linewidth=1pt](4,0.4){0.4}}
\rput(0,0){\pscircle[linewidth=1pt,fillstyle=solid, fillcolor=black](4,0.4){0.03}}
\rput(4.55,0.5){\makebox(0,0)[lc]{$\cdots$}}
\rput(0,0){\pscircle[linewidth=1pt](5.7,0.6){0.6}}
\rput(0,0){\pscircle[linewidth=1pt,fillstyle=solid, fillcolor=black](5.7,0.6){0.03}}
\rput(0,0){\pscircle[linewidth=1pt](7,0.7){0.7}}
\rput(0,0){\pscircle[linewidth=1pt,fillstyle=solid, fillcolor=black](7,0.7){0.03}}
%
\rput(0,0){\psline[linewidth=1pt,linecolor=red](0,1.4)(7,1.4)}
\rput(0,0){\psarc[linewidth=1pt,linecolor=red](0,0.7){0.7}{-10}{90}}
\rput(0,0){\psarc[linewidth=1pt,linecolor=red](1.2,0.5){0.5}{0}{170}}
\rput(0,0){\psarc[linewidth=1pt,linecolor=red](2.2,0.5){0.5}{20}{180}}
\rput(0,0){\psarc[linewidth=1pt,linecolor=red](4,0.4){0.4}{33}{163}}
\rput(0,0){\psarc[linewidth=1pt,linecolor=red](5.7,0.6){0.6}{5}{180}}
\rput(0,0){\psarc[linewidth=1pt,linecolor=red](7,0.7){0.7}{90}{185}}
%
%
\rput(0,0){\psline[linewidth=0.5pt](0,0.7)(1.2,0.5)}
\rput(0,0){\psline[linewidth=0.5pt](2.2,0.5)(1.2,0.5)}
\rput(0,0){\psline[linewidth=0.5pt](2.2,0.5)(2.7,0.7)}
\rput(0,0){\psline[linewidth=0.5pt](4,0.4)(4.45,0.7)}
\rput(0,0){\psline[linewidth=0.5pt](4,0.4)(3.5,0.55)}
\rput(0,0){\psline[linewidth=0.5pt](5.7,0.6)(5,0.6)}
\rput(0,0){\psline[linewidth=0.5pt](5.7,0.6)(7,0.7)}
\rput(0,0){\psline[linewidth=0.5pt](0,0.7)(0,0)}
\rput(0,0){\psline[linewidth=0.5pt](1.2,0)(1.2,0.5)}
\rput(0,0){\psline[linewidth=0.5pt](2.2,0.5)(2.2,0)}
\rput(0,0){\psline[linewidth=0.5pt](4,0.4)(4,0)}
\rput(0,0){\psline[linewidth=0.5pt](5.7,0.6)(5.7,0)}
\rput(0,0){\psline[linewidth=0.5pt](7,0)(7,0.7)}

\rput(0,-0.05){\makebox(0,0)[ct]{\tiny$z_1{=}0$}}
\rput(1.2,-0.05){\makebox(0,0)[ct]{\tiny$z_2$}}
\rput(2.2,-0.05){\makebox(0,0)[ct]{\tiny$z_3$}}
\rput(2.85,-0.05){\makebox(0,0)[lt]{$\cdots$}}
\rput(4,-0.05){\makebox(0,0)[ct]{\tiny$z_i$}}
\rput(4.55,-0.05){\makebox(0,0)[lt]{$\cdots$}}
\rput(5.7,-0.05){\makebox(0,0)[ct]{\tiny$z_{n{-}2}$}}
\rput(7,-0.05){\makebox(0,0)[ct]{\tiny$z_{n{-}1}=1$}}
%
\rput(0.05,0.3){\makebox(0,0)[lb]{\tiny$r_{1}$}}
\rput(1.25,0.2){\makebox(0,0)[lb]{\tiny$r_{2}$}}
\rput(2.25,0.2){\makebox(0,0)[lb]{\tiny$r_{3}$}}
\rput(4.05,0.15){\makebox(0,0)[lb]{\tiny$r_{i}$}}
\rput(5.75,0.3){\makebox(0,0)[lb]{\tiny$r_{n{-}2}$}}
\rput(6.95,0.35){\makebox(0,0)[rb]{\tiny$r_{n{-}1}$}}
}
}
\end{pspicture}
$$
Note that $r_1=r_{n-1}$. Using hyperbolic geometry, it is not difficult to show that
$$
\lambda_{1,i}=\frac{z_i}{2\sqrt{r_1r_i}}\quad\hbox{for}\quad 2\le i\le n-1
$$
and
$$
\lambda_{0,i}=\sqrt{r_1/r_i} \quad\hbox{for}\quad 1\le i\le n-1,
$$
so
\be
x_i=\frac{z_i}{2r_1} \quad \hbox{for}\quad 2\le i\le n-1,\qquad \hbox{and}\qquad \widehat \xi_i = \frac{z_i-z_{i-1}}{z_2}\quad \hbox{for}\quad i\le 3\le n-1.
\ee

The integration measure then have a form similar to that in the ``fool's crown'' case:
\be\label{measure-disc}
d\Omega=\frac{\mathop{\wedge}\limits_{i=2}^{n-2} dz_i }{z_2(z_3-z_2)(z_4-z_3)\cdots (z_{n-2}-z_{n-3})(1-z_{n-2})}.
\ee

We now turn to the action. We have that
$$
s_1=\sqrt{\frac{r_1}{r_2}},\quad s_i=\sqrt{\frac{r_i}{r_{i-1}}}+\sqrt{\frac{r_i}{r_{i+1}}},\ 2\le i\le n-2,\quad s_{n-1}=\sqrt{\frac{r_{n-1}}{r_{n-2}}},\quad r_0=\frac1{2r_1}.
$$
Using that
$$
(z_i-z_{i-1})=2\sqrt{r_ir_{i-1}}\quad \hbox{for}\quad 2\le i\le n-1,
$$
for the action we obtain (taking into account that $r_{n-1}=r_1$)
\begin{align}
S&=\log \sqrt{\frac{r_1}{r_2}} +\sum_{i=2}^{n-2} \log\left( \sqrt{\frac{r_i}{r_{i-1}}}+\sqrt{\frac{r_i}{r_{i+1}}}\right) +\log \sqrt{\frac{r_{n-1}}{r_{n-2}}} +\log\frac{1}{2r_1}\nonumber\\
&=\sum_{i=2}^{n-2} \log\Bigl(\sqrt{r_ir_{i-1}}+\sqrt{r_ir_{i+1}} \Bigr)  - \log\Bigl(2 r_1^{1/2} r_2\cdots r_{n-2} r_{n-1}^{1/2} \Bigr)\nonumber\\
&=\log\Bigl[ z_3(z_4-z_2)\cdots (z_{n-2}-z_{n-4})(1-z_{n-3}) \Bigr] \nonumber\\
&\qquad\qquad\qquad - \log\Bigl[ z_2(z_3-z_2)(z_4-z_3)\cdots (z_{n-2}-z_{n-3})(1-z_{n-2}) \Bigr].
\label{S-disc}
\end{align}
We again observe that all ``unwanted'' terms in $d\Omega$ resulting in singularities are cancelled with the terms in $e^{-S}$, and we have the following proposition, which remains valid also in the case of even $n$.
\begin{proposition}\label{pr:volumes-disc}
Let $0<\delta_i=z_i-z_{i-1}$, $2\le i\le n-1$. The volumes $V^{\text{disc}}_{n}$ for the moduli spaces of disc with $n\ge 4$ boundary cusps are given by the integrals
$$
V^{\text{disc}}_{n}=\int d\Omega\,e^{-S}=\int_{\delta_2+\cdots +\delta_{n-2}\le 1} \frac{ \prod_{i=2}^{n-2}d\delta_i }{\prod_{i=2}^{n-2} \bigl(\delta_i+\delta_{i+1} \bigr)}.
$$
with $\delta_2+\delta_3+\cdots+\delta_{n-2}+\delta_{n-1}=1$ These volumes are finite for all $n$.
\end{proposition}

\subsection{Examples}
\subsubsection{$n=5$}
This is the first nontrivial example. We have
$$
V^{\text{disc}}_{5} = \int_0^1 dz_3\int_0^{z_3} dz_2\,\frac{1}{z_3(1-z_2)}=\int_0^1 \frac{dz_3}{z_3}(-\log(1-z_3))=Li_2(1)=\frac{\pi^2}{6}.
$$

\subsubsection{$n=6$}
Here we again use the existence of the flat direction: for $n=2k$ the vector field $\sum_{i=1}^{k-1}\partial_{z_{2i}}$ leaves the integrand invariant, and the integral reads
\begin{align*}
&\int_0^1 \frac{dz_3}{z_3(1-z_3)}\int_0^{z_3}dz_2\int_{z_3}^1 dz_4 \frac 1{z_4-z_2}\\
=&\int_0^1 \frac{dz_3}{z_3(1-z_3)} \bigl[ -z_3\log(z_3)-(1-z_3)\log(1-z_3) \bigr] \\
=&\int_0^1 dz_3 \Bigl[ -\frac{\log z_3}{1-z_3}-\frac 1 {z_3} \log(1-z_3) \Bigr] \\
=&2 \int_0^1 dz_3 \Bigl[ -\frac 1 {z_3} \log(1-z_3) \Bigr]=2Li(1)=\frac{\pi^2}{3}.
\end{align*}

\subsection{The continuum limit}

The continuum limit of the action, assuming $z_i\to f(t_i=i/n)$, where $f:[0,1]\to [0,1]$ is smooth with $f'(x)>0$, after canceling singularities of the measure is merely
\begin{align}
&\sum_i \log(z_{i+1}-z_{i-1})=\sum_i \log\left(f'(t_i)2\varepsilon + \frac 16 f'''(t_i) 2\varepsilon^3 +\dots\right)\nonumber\\
&=\sum_i \left( \log(f'(t_i)) +\frac{\varepsilon^2}{6}\frac{f'''(t_i)}{f'(t_i)}\right)\sim \int_0^1 dt\,\left( \frac{1}{\varepsilon} \log(f'(t)) +\frac{\varepsilon}{6}\,\frac{f'''(t)}{f'(t)}\right)=\frac{1}{\varepsilon} A_1^{(0)}+\frac{\varepsilon}{6}\int_0^1 dt\, S[f,t].
\end{align}
We therefore do not have the term with $f'(t)^2$ in this case if we identify the variable $t$ with the continuum limit of $z_i$. Note that if we use exponentiated quantities, e.g., $z_i=e^{\zeta_i}$, then such a term will appear due to the cocycle property of the Schwarzian. So, it is really scheme-dependent.

For the continuum limit of the symplectic form, by the same reason, we just obtain the pure Gelfand--Fuchs cocycle term (\ref{GF}).

\section{Concluding remarks}
\setcounter{equation}{0}
In conclusion we collect open questions and possible directions of development of structures described in the main text.
\begin{itemize}
\item It is tempting to investigate in full asymptotic behavior of the volumes $V_{n,P}^{\text{crown}}$ and $V_{n}^{\text{disc}}$. We expect a relation to resurgence procedure developed by Eynard et al in \cite{EGGLS}. 
\item A related problem is to develop (quasi)recurrence relations for the above integrals and look for a connection to Topological Recursion theory.
\item Structures of crowns may be interesting from the complex-analytic point of view: we can interpret the crown as a set of monodromies of a system with a very high-order pole; such structures played a pivotal role in the description of monodromies of Painlev\'e equations in \cite{CMR2}.
\item Action functionals in Theorem~\ref{th:volumes} and Proposition~\ref{pr:volumes-disc} describe in physical terms interesting systems of $n$ particles on a circle (in the ``fool's crown'' case) and on a line segment (in the case of a disc) with a Coulomb-like interaction $\sim \log(z_i-z_{i-2})$ between next-to-nearest neighbors and rigid-wall interaction between nearest neighbors. It is interesting to address physical properties of such systems.
\item It might be that the above bi-partition of sites of such linear lattices can be developed into description of spin-$1/2$ systems. A far-reaching generalization can be 2D lattice structures appearing in cluster algebra description of the symplectic groupoid developed by the author and M. Shapiro in \cite{ChSh2}.
\item It might be interesting to explore action functionals introduced in this text as Yang--Yang functionals studied in \cite{NRS} for $\Sigma_{g,s}$. 
\item We are going to explore in more details the dynamics imposed by the action (\ref{action-shear}) on the set of shear coordinates $y_i$. In particular, in the case of even $n$, we can split the set of $y_i$ into the subsets of odd $y_{2j+1}$ and even $y_{2j}$ subsets, identifying each of these subsets with coordinates and momenta. Note that, in this case, the total coordinate $X:=\sum_j y_{2j+1}$ and momentum $P:=\sum_j y_{2j}$ are preserved being Casimirs of this system. 
\item We are about to explore the continuum limit as an asymptotic expansion w.r.t. the parameter $\varepsilon$. Here, the first question is related to the main result of the first Stanford and Witten paper \cite{SW1}: is our action (\ref{act1}) ``one-loop exact'' in any sense? Argumentation of the authors of \cite{SW1} was based on ``hidden'' sypersymmetry of their action stemmed from its fermionic realization in SYK model.
\item Among many possible extensions of our approach let us mention, first, a supersymmetrization of the model based on JT gravity approach of \cite{SW} combined with a version of a supersymmetric cluster algebras developed by Penner and Zeitlin \cite{Penner-Zeitlin}. A closely related topic is fermionization of our structures regarding their possible relations to the SYK model. 
\item Finally, let us mention a possible extension of our action to the case of nonorientable trumpets developed by Stanford and Witten in \cite{SW}. 
\end{itemize}
All this indicates that, hopefully, this model has some potential of development.

\section*{Acknowledgements}
I am grateful to Volodya Rubtsov and Misha Shapiro for encouragement and useful discussion. Special thanks are due to Anton Alexeev for careful reading of the manuscript and numerous valuable remarks and suggestions. 

\setcounter{section}{0}
\appendix{Asymptotic regimes for $V_{n,P}^{\text{crown}}$.}\label{AppA}

\subsection{Regime $P\to\infty$.} 
We first consider the limit as $P\to\infty$ of the volumes $V_{n,P}^{\text{crown}}$ from Theorem~\ref{th:volumes}. 

Let us scale $\delta_i=Pa_i$ with $\sum_{i=1}^n a_i=1$. The volume is then given by the integral
$$
V_{n,P}^{\text{crown}}=\frac{1}{2^{n-1}}P^{n-1}\int_{a_1+a_2+\cdots+a_{n-1}\le 1}da_1\,da_2\cdots da_{n-1}\frac{e^P-1}{\prod_{i=1}^n (e^{P(a_i+a_{i+1})}-1)}.
$$
We first consider the integral ``in the bulk'' with $Pa_i$ large positive numbers. We can then neglect $-1$'s in denominators and just obtain using that $\sum a_i=1$,
\be\label{Pinfin}
\frac{1}{2^{n-1}}P^{n-1}\int_{a_1+a_2+\cdots+a_{n-1}\le 1}da_1\,da_2\cdots da_{n-1}\frac{e^P}{\prod_{i=1}^n e^{P(a_i+a_{i+1})}}=
\frac{1}{2^{n-1}}P^{n-1} \frac{1}{(n-1)!} e^{-P}.
\ee

We next consider the same integral in the regime as $a_1+a_2\le 1/P$. Then, since 
$$
\frac{1}{e^{P(a_1+a_2)}}< \frac{1}{P(a_1+a_2)},
$$
for the integral over $a_1$ and $a_2$, we have (we also use that $e^P-1\sim e^P$ and that other variables are ``in the bulk'')
\begin{align*}
&\frac{1}{2^{n-1}}P^{n-1} \int_0^{1/P} d(a_1+a_2) \int_{0}^{a_1+a_2} da_1 \frac{1}{P(a_1+a_2)} \int_{(a_1+a_2)+a_3+\cdots+a_{n-1}\le 1}  \frac{ da_3\,\dots da_{n-1}\, e^P}{ \prod_{i=2}^n (e^{P(a_i+a_{i+1})}-1)}\\
=&\frac{1}{2^{n-1}}P^{n-1} \int_0^{1/P} d(a_1+a_2)  \int_{(a_1+a_2)+a_3+\cdots+a_{n-1}\le 1} da_3\,\dots da_{n-1} \frac{(a_1+a_2)}{P(a_1+a_2)} \frac{e^P}{e^{2P-a_1-a_2}} \\
=&\frac{P^{n-1}}{2^{n-1}}\,  \frac{1}{P e^P} \int_0^{1/P} d(a_1+a_2) e^{P(a_1+a_2)} \int_{(a_1+a_2)+a_3+\cdots+a_{n-1}\le 1}  da_3\,\dots da_{n-1} \\
\sim & \frac{P^{n-2}e^{-P}}{2^{n-1}} (e^{1}-1)\frac{1}{(n-3)!}\sim \hbox{Const} \, P^{n-2}e^{-P},
\end{align*}
so this integration is of order $O(1/P)$ w.r.t. the ``main'' integration in the bulk (\ref{Pinfin}). We therefore conclude that 
\be\label{VolInfin}
V_{n,P}^{\text{crown}}=\frac{P^{n-1}e^{-P}}{2^{n-1}(n-1)!}(1+O(1/P))\quad \hbox{as}\quad P\to\infty.
\ee
In particular, for $n=3$, we have $\lim_{P\to\infty} V_{3,P}^{\text{crown}}=\frac18 P^2 e^{-P}$, in agreement with  (\ref{V3P}).

\subsection{Regime $P\to 0$.} In this case, we just replace $e^{\delta_i+\delta_{i+1}}-1$ by $\delta_i+\delta_{i+1}$. for the volume we then just have
$$
\frac{1}{2^{n-1}}\int_{\delta_1+\dots+\delta_{n-1}\le P}\frac{ d\delta_1\,\cdots d\delta_{n-1} \cdot P}{\prod_{i=1}^n(\delta_i+\delta_{i+1})},
$$
and using that $P=\delta_1+\cdots+\delta_n$ and obvious cyclic symmetry of the integral, we can cancel one factor (say, $\delta_n+\delta_1$) in the denominator thus obtaining
\begin{align*}
&\frac{1}{2^{n-1}}\cdot \frac{n}{2} \int_{\delta_1+\dots+\delta_{n}= P} \frac{ d\delta_1\,\cdots d\delta_{n-1}}{(\delta_1+\delta_2)(\delta_2+\delta_3)\cdots (\delta_{n-1}+\delta_n)}\\
&\qquad\qquad\qquad = \frac{n}{2^{n}} \int_{a_1+\dots+a_{n}= 1} \frac{ da_1\,\cdots da_{n-1}}{(a_1+a_2)(a_2+a_3)\cdots (a_{n-1}+a_n)}.
\end{align*}
We therefore have that
\be
\lim_{P\to 0} V_{n,P}^{\text{crown}} =  \frac{n}{2^{n}} q_n,\quad \hbox{where}\quad q_n=\int_{a_1+\dots+a_{n}= 1} \frac{ da_1\,\cdots da_{n-1}}{(a_1+a_2)(a_2+a_3)\cdots (a_{n-1}+a_n)}.
\ee
We now evaluate $q_3$ and $q_4$.
\subsubsection{Evaluating $q_3$.}
\begin{align*}
q_3&=\int_{0}^1 da_2\int_0^{1-a_2}\frac {da_1}{(a_1+a_2)(1-a_1)}=\int_0^1 \frac{da_2}{a_2+1}\log(1/a_2^2)\\
&=-2\int_0^1 \frac{da_2}{1+a_2}\log(a_2)=-2\int_0^1 d\log(a_2+1) \cdot\log(a_2)=2\int_0^1\frac{\log(1+a_2)}{a_2}da_2=2\cdot \frac{\pi^2}{12}=\frac{\pi^2}{6}
\end{align*}
We therefore have that
$$
\lim_{P\to 0} V_{3,P}^{\text{crown}} = \frac{3}{8} q_3=\frac{\pi^2}{16},
$$
in full agreement with (\ref{V3P}).

\subsubsection{Evaluating $q_4$.}
\begin{align*}
q_4&=\int_0^{1} da_3\int_0^{1-a_3}da_2\int_0^{1-a_2-a_3}\frac{da_1}{(a_1+a_2)(a_2+a_3)(1-a_1-a_2)}\\
&=\int_{a_2+a_3\le 1} \frac{da_3\,da_2}{a_2+a_3} \log\Bigl[ \frac{1-a_3}{a_3}\cdot \frac{1-a_2}{a_2} \Bigr] \\
&= \int_{a_2+a_3\le 1}  \frac{da_3\,da_2}{a_2+a_3} \Bigl[ \log\frac{1-a_3}{a_3} +\log \frac{1-a_2}{a_2} \Bigr]\\
&= \int_{a_2+a_3\le 1}  \frac{da_3\,da_2}{a_2+a_3} \Bigl[ \log\frac{1-a_3}{a_3} +\log \frac{1-a_2}{a_2} \Bigr]\\
&=\frac12 \int_0^{1} da_3  \log\frac{1-a_3}{a_3} \int_0^{1-a_3} \frac{da_2}{a_2+a_3}\\
&=\frac12 \int_0^{1} da_3  \log\frac{1-a_3}{a_3}\cdot \log(1/a_3)\\
\end{align*}
So, we have 
\begin{align*}
q_4&=\frac12 \int_0^1 dx\,\log\frac{1-x}{x}\cdot \log\frac{1}{x}= \int_0^1 d\bigl( (x-1)\log(1-x) -x\log x\bigr)\cdot\log \frac{1}{x}\\
&=\frac12 \int_0^1 dx \bigl( (x-1)\log(1-x) -x\log x\bigr)\cdot \frac{1}{x}\\
&=-\frac12 \int_0^1 dx \frac1x \log(1-x)=\frac12 Li(1)=\frac{\pi^2}{12}.
\end{align*}
We therefore have 
$$
\lim_{P\to 0} V_{4,P}^{\text{crown}} = \frac{4}{16} q_4=\frac{\pi^2}{48},
$$

\appendix{Changing the variables in the form $\omega$ (\ref{GF})}
\label{AppB}
\setcounter{section}{0}
Let $g(t)=g(f(t))$ in (\ref{GF}). We then have (taking into account an obvious skew-symmetricity of the expression and that $df(x)\wedge df(x)=0$ and  $df'(t)\wedge df'(t)=0$)
\begin{align*}
&-\frac14 \int dt\, d\bigl( \log f'(t)+\log g'(f(t))\bigr)\wedge \partial_t\, d \bigl( \log f'(t)+\log g'(f(t))\bigr)\\
=&-\frac14 \int dt\, d\bigl( \log f'(t)\bigr)\wedge \partial_t\, d \bigl( \log f'(t)\bigr)-\frac12 \int dt\, d\bigl( \log f'(t)\bigr)\wedge \partial_t\, \Bigl( \frac{g''(f(t))}{g'(f(t))} df(t)\Bigr) \\ &\qquad\qquad\qquad\qquad
-\frac14 \int dt\, \frac{g''(f(t))}{g'(f(t))} df(t)\wedge \partial_t\, \Bigl( \frac{g''(f(t))}{g'(f(t))} df(t)\Bigr)\\
=&-\frac14 \int dt\,  d\bigl( \log f'(t)\bigr)\wedge \partial_t\, d \bigl( \log f'(t)\bigr)-\frac12 \int dt\, \frac{df'(t)}{f'(t)}\wedge \partial_t\, \Bigl( \frac{g''(f(t))}{g'(f(t))} \Bigr)df(t) \\ &\qquad\qquad\qquad\qquad
-\frac14 \int dt\, \frac{g''(f(t))}{g'(f(t))} df(t)\wedge \Bigl( \frac{g''(f(t))}{g'(f(t))} df'(t)\Bigr)\\
=&-\frac14 \int dt\, d\bigl( \log f'(t)\bigr)\wedge \partial_t\, d \bigl( \log f'(t)\bigr)-\frac12 \int dt\, \frac{df'(t)}{f'(t)} \wedge df(t)  \Bigl( \frac{g'''(f(t))}{g'(f(t))}-\frac{\bigl(g''(f(t))\bigr)^2}{\bigl(g'(f(t))\bigr)^2} \Bigr)\cdot f'(t) \\ &\qquad\qquad\qquad\qquad
-\frac14 \int dt\, df(t)\wedge df'(t) \frac{\bigl(g''(f(t))\bigr)^2}{\bigl(g'(f(t))\bigr)^2}\\
=&-\frac14 \int dt\, d\bigl( \log f'(t)\bigr)\wedge \partial_t\, d \bigl( \log f'(t)\bigr)-\frac12 \int dt\, df'(t) \wedge df(t)\Bigl( \frac{g'''(f(t))}{g'(f(t))}
-\frac32 \frac{\bigl(g''(f(t))\bigr)^2}{\bigl(g'(f(t))\bigr)^2} \Bigr),
\end{align*}
that is,
\begin{align}\label{g-f}
&-\frac 14 \int_{0}^{1} dt\, d\bigl[ \log\bigl( \partial_t[ g(f(t))]\bigr)  \bigr] \wedge d \left( \partial_t \bigl[ \log\bigl(  \partial_t [g(f(t))]\bigr)  \bigr] \right)\nonumber \\
&=-\frac14 \int dt\, d\bigl( \log f'(t)\bigr)\wedge \partial_t\, d \bigl( \log f'(t)\bigr)-\frac12 \int dt\, df'(t) \wedge df(t) S[g,f],
\end{align}
where $S[g,f]:=  \frac{g'''(f)}{g'(f)}-\frac32 \frac{(g''(f))^2}{(g'(f))^2}$ is the Schwarzian derivative (\ref{Schw}) of the function $g(f)$.

\end{document}